\documentclass[a4paper,11pt]{article}
\pdfoutput=1 
\usepackage{jcappub}
\usepackage{graphicx}
\usepackage{bm}
\pdfoutput=1 
\usepackage{color}
\usepackage{makecell}
\usepackage{afterpage}
\usepackage[dvipsnames]{xcolor}
\usepackage{multirow}
\usepackage{cleveref}
\usepackage{subfigure}
\usepackage{booktabs}
\usepackage[normalem]{ulem}		
\usepackage{pgfgantt}

\newcommand{\ii}{{\rm{i}}}
\newcommand{\fett}[1]{{\boldsymbol{#1}}}

\newcommand{\be}{\begin{equation}}
\newcommand{\ee}{\end{equation}}

\newcommand{\kop}
{\mathfrak{K}}

\newcommand{\Hc}{\mathcal{H}}

\newcommand{\CLASS}{{\sc class}}
\newcommand{\COSIRA}{{\sc cosira}}

\newcommand{\HL}{H_{\rm L}}
\newcommand{\HT}{H_{\rm T}}
\newcommand{\nab}{\nabla}

\title{Fast simulations of cosmic large-scale structure with massive neutrinos}

\author[1]{Christian Partmann,}

\emailAdd{christian.partmann@rwth-aachen.de}

\author[1]{Christian Fidler,}

\emailAdd{fidler@physik.rwth-aachen.de}

\author[2]{Cornelius Rampf}

\emailAdd{cornelius.rampf@oca.eu}

\author[2]{and Oliver Hahn}

\emailAdd{oliver.hahn@oca.eu}

\affiliation[1]{Institute for Theoretical Particle Physics and Cosmology (TTK), \\ RWTH Aachen University, D-52056 Aachen, Germany.}

\affiliation[2]{Laboratoire Lagrange, Universit\'e C\^ote d'Azur, Observatoire de la C\^ote d'Azur, CNRS, Blvd de l'Observatoire,
CS 34229, 06304 Nice cedex 4, France.}

\abstract{
Accurate cosmological simulations that include the effect of non-linear matter clustering as well as of massive neutrinos are essential for measuring the neutrino mass scale from upcoming galaxy surveys. Typically, Newtonian simulations are employed and the neutrino distribution is sampled with a large number of particles in order to beat down the shot noise. Here we perform very efficient simulations with light massive neutrinos which require virtually no extra cost over matter-only simulations, and furthermore do not require tracer particles for the neutrinos. Instead, we use a weak-field dictionary based on the recently developed Newtonian motion approach, where Newtonian simulations for matter are paired with a linear relativistic Boltzmann code to allow for an absorption of the neutrino evolution into a time-dependent coordinate transformation. For this, only minimal modifications in existing N-body codes are required, which we have explicitly implemented in \emph{gevolution} and {\sc gadget-2}. Our fast method determines the non-linear matter power spectrum to permille-level precision when compared against state-of-the-art simulations that have been performed for $0.1\,{\rm eV} \leq \sum m_\nu \leq 0.3$\,eV. 
}

\begin{document}

\hfill{\small TTK-20-06}

\maketitle

\section{Introduction}

The next generation of galaxy surveys such as EUCLID \cite{euclid} and LSST \cite{lsst} will resolve the large-scale structure to unprecedented accuracy.
Such data will allow us to reduce uncertainties of the parameters of the standard cosmological model, the $\Lambda$CDM model, to the sub-percent level. Thus, it is crucial to theoretically determine the large-scale structure to sufficient accuracy~\cite{Hearin_2012}. 

In contrast to cold matter, massive neutrinos possess large thermal velocities which hinder them to cluster non-linearly on scales smaller than their free-streaming length~\cite{Lesgourgues:2012uu}. The free-streaming length is (roughly) inversely proportional to the neutrino mass scale, implying that the lighter the neutrinos the lesser they participate in small-scale clustering. For this reason, the large-scale evolution of neutrinos is well described in linear theory if the neutrino masses are sufficiently small (which appears to be the case; see below). Furthermore, neutrino masses can have an effect on the time of matter-radiation equality (depending which parameters are fixed/varied in the analysis) which leads to a pronounced suppression suppression of the total matter power spectrum on small scales. Due to the measurable imprint on the matter power spectrum, upcoming large-scale structure surveys are thus in a very promising position to determine the absolute neutrino mass scale~\cite{Jimenez:2016ckl}.

To keep up with the increasing sensitivity of large-scale structure observations, 
several methods for the inclusion of massive neutrinos in cosmological N-body simulations have been proposed. 
Using significant computational effort, some methods explicitly sample the neutrino phase-space with particles (e.g. \cite{Brandbyge:2008rv, 2011MNRAS.410.1647A, 2012MNRAS.420.2551B, Villaescusa-Navarro:2013pva, Carbone:2016nzj,Brandbyge_2019, Adamek:2017uiq, Banerjee:2018bxy, Emberson:2016ecv}) which allows for a fully non-linear treatment of neutrino perturbations. Such a fully non-linear modeling of neutrino evolution is particularly important for heavy neutrinos, that can e.g. be appreciably accreted by the massive galaxy clusters \cite{Ichiki_2012,2014PhRvD..89f3502L,2014PhRvD..90h3518L}. A heavy neutrino scenario is however increasingly disfavoured by cosmological observations~\cite{Aghanim:2018eyx}. Furthermore, this fully non-linear approach comes with the serious limitation that a very large number of neutrino particles is required, $N_\nu \gtrsim \mathcal{O}(10^{11})$,  in order to reduce the shot noise to a level that is sufficient to measure effects in the non-linear neutrino power spectrum \cite{Adamek:2017uiq}. See however~\cite{Banerjee:2018bxy} for recent investigations to reduce shot noise by coarse-graining the initial neutrino momentum distribution together with applying a regular sampling method (as opposed to random sampling) of the neutrino phase-space.

Flavour oscillation experiments provide a lower bound on the difference of neutrino masses (which implies $\sum m_{\nu} \gtrsim 0.06$\,eV for the normal masss hierarchy and $\sum m_{\nu} \gtrsim 0.1$\,eV for the inverted hierarchy). By contrast,  current measurements of the cosmic microwave background anisotropies from PLANCK in combination with data from the baryonic acoustic oscillations provide an upper bound to the sum of the neutrino masses, $\sum m_\nu < 0.12$\,eV at 95\% confidence~\cite{Vagnozzi_2017, Aghanim:2018eyx}.
For such light neutrinos, the computationally expensive non-linear treatment of neutrinos can be avoided by assuming that perturbations remain close to linear at all times.
The grid-based approaches of~\cite{Brandbyge:2008js,Brandbyge:2016raj,Tram:2018znz}  explicitly exploit this, by sampling the linear neutrino perturbations on grid points which are added to the Poisson solver of the N-body simulation. A similar grid-like method employing instead a linear response function approach has been introduced in~\cite{2013MNRAS.428.3375A} (see also~\cite{Liu:2017now}). There exist also some hybrid approaches where the aforementioned approaches are combined with a particle-based method, such as \cite{Brandbyge:2009ce,Bird:2018all} that evolve neutrinos on grid points and subsequently convert them into particles (at a predefined momentum-to-temperature ratio), in order to allow for non-linear neutrino clustering. Another technique has been recently reported in~\cite{Zennaro:2019aoi}, which incorporates the effect of linear massive neutrinos by means of a rescaling procedure (itself based on the marriage of the methods of~\cite{Zennaro:2016nqo} and~\cite{Angulo:onetofitthemall}). Finally, for light neutrinos where a linear treatment suffices, also analytic approximation schemes using perturbation theory are amenable, see e.g.\ \cite{Wong:2008ws,Blas:2014hya,Fuhrer:2014zka,Senatore:2017hyk,deBelsunce:2018xtd, Saito_2008}.

Another challenge for the computation of large-scale structure observables is to account for general relativistic effects in a consistent way. It is well known that the output of standard Newtonian N-body simulations does not directly match the results of general relativity (GR) on large scales \cite{Chisari:2011iq,Fidler:2015npa,Hahn:2016roq}. However, the recently developed weak-field approach \cite{Fidler:2017pnb}  that accounts for leading-order general relativistic effects as well as non-linear clustering on small scales provides a relativistic interpretation of standard N-body simulations. It was shown, that N-body simulations containing only one cold species can be made compatible with general relativity if their output is interpreted in a tailor-made gauge \cite{Fidler:2016tir}, leading to the idea of Newtonian motion gauges (see the following section for a concise summary, or appendix~\ref{app:Nmgauge} for more technical details).

In this paper, we run Newtonian simulations exploiting the Newtonian motion gauge approach, thereby allowing us to include massive neutrinos in terms of a post-processing, while delivering a solution that is in accordance with the weak-field limit of general relativity. The specific technical setup necessary for this has been developed especially by the references~\cite{Fidler:2018bkg} and \cite{Fidler:2018stc}. We summarise the general idea of the approach in the following section, while more specific details needed for the implementation are outlined in section~\ref{sec:Method}. Numerical results of our method are presented in section~\ref{sec:results}, while 
further results are given in section~\ref{sec:other_stuff}. Specifically, an alternative implementation of the approach is outlined in  section~\ref{alternativeimpl}, and the essential steps for performing a lightcone analysis are outlined in section~\ref{sec:lightcone}. We conclude in~\ref{sec:conclusions} where we also outline some future work.

\section{Simulations and the Newtonian motion (Nm) gauge approach}\label{sec:recap}

Standard Newtonian N-body simulations that evolve  matter essentially solve the Newtonian equation of motion to obtain the family of matter trajectories $\fett{x}^{\rm N}$ 
\be \label{EoM-N}
  \partial_t^2 \fett{x}^{\rm N} + 2 H \partial_t \fett{x}^{\rm N} = - \nabla \, \Phi^{\rm N}(\fett{x}^{\rm N})  \,,\, \qquad \qquad \quad \rm (Newtonian~simulation) 
\ee
where $t$ denotes the (convective) time derivative,
 $H(t) = \dot a/a$ is the Hubble parameter where $a$ is the cosmic scale factor, and~$\Phi^{\rm N}$ is the Newtonian gravitational potential that is sourced by the non-linear matter density. Here and in the following, the superscript 'N' denotes quantities as determined within a Newtonian setup.

General relativity and the presence of additional species, such as massive neutrinos, distort the Newtonian matter trajectories. The combined impact can be described by adding a yet unspecified force correction 
$\delta\fett{F}(\fett{x}, t)$ to the formerly Newtonian equation of motion,
\be \label{relatEOM}
    \partial_{t}^2 \fett{x} + 2 H \partial_t \fett{x} = - \nabla \, \Phi^{\rm N}(\fett{x}) + \delta\fett{F}(\fett{x},t) \,.  \qquad\quad \qquad \rm  (relativistic~trajectories)
\ee
Clearly, the Newtonian and relativistic trajectories given respectively by Eqs.\,\eqref{EoM-N} and~\eqref{relatEOM} are in general not identical. The function $\delta\fett{F}(\fett{x}, t)$ contains terms coming from other fluid species, metric perturbations and anisotropic stress (for an explicit expression see Eq.\,(4.12) in~\cite{Fidler:2017pnb} where it is called $\fett{\gamma}$).
In general relativity, we have the gauge freedom to choose the underlying coordinate system, which allows us to absorb dynamical effects along a trajectory into the definition of the employed coordinates. This is the essential idea of the Newtonian motion (Nm) gauge approach. We thus choose a set of new coordinates, employing the coordinate shift $\fett{L}$,
\be \label{coordinateshift}
  \fett{x}^{\rm Nm} = \fett{x} + \fett{L}  \,,  \,\,\qquad \quad \qquad\qquad \qquad\quad \rm (Nm~coordinates)
\ee 
defined by the condition that in these coordinates, we recover precisely the (Newtonian) equation of motion
\be \label{EoM-Nm}
  \partial_t^2 \fett{x}^{\rm Nm} + 2 H \partial_t \fett{x}^{\rm Nm} = - \nabla \Phi^{\rm N}(\fett{x}^{\rm Nm}) \,, \qquad\qquad\qquad \qquad \rm  (relativistic~Nm~trajectories)
\ee 
where again the potential $\Phi^{\rm N}$ is only determined by the distribution of
matter (within our new coordinates).

In the Nm approach, incorporating massive neutrinos amounts to solving for the coordinate shift $\fett{L}$. 
 Indeed, plugging~\eqref{coordinateshift} into~\eqref{EoM-Nm}, one obtains upon identification with~\eqref{relatEOM} a second-order differential equation in time for~$\fett{L}$. This differential equation requires two initial conditions (ICs), for example $\fett{L} \rvert_{t_{\rm ini}}$ and $\partial_t \fett{L}\rvert_{t_{\rm ini}}$, which are prescribed at initial time $t_{\rm ini}$ (for the approach employed in sections~\ref{sec:Method}--\ref{sec:results}, $t_{\rm ini}$ is the time when the simulation is initialised).
 Although many choices of initial data are feasible, we choose that $\fett{L} \rvert_{t_{\rm ini}}=0= \partial_t \fett{L} \rvert_{t_{\rm ini}} $ and thus, $\fett{x}^{\rm Nm}$ and $\fett{x}$ coincide at the initial time. Then, the shift $\fett{L}$ can be understood as the accumulated effect that the extra species and relativistic corrections have on the dynamics of the matter particles.

In the relativistic context, $\fett{L}$ is the spatial gauge generator that is required to achieve formal agreement between the Newtonian equations of motion for 
matter with the relativistic ones in the weak-field limit. This gauge generator can be expressed in terms of the scalar metric potential $\HT$, which is the off-diagonal part of the spatial metric (for a definition see Eq.\,\eqref{metric-potentials}). The weak-field limit only assumes that the metric potentials remain perturbatively small, while the fluid variables for matter are allowed to grow arbitrarily large. The formalism is therefore well suited to bridge the gap between non-linear simulations and cosmological perturbation theory. 

While the spatial gauge is determined by the Nm gauge condition, the temporal gauge fixing can be done almost freely within the Nm approach. Indeed there exists a class of suitable time coordinates for which the underlying weak-field assumptions are not violated, and a good choice is the time coordinate of the Poisson gauge (commonly also called ``Newtonian gauge'') --- which we choose also in the following. For further technical details about the underlying weak-field assumptions together with the derivations of relevant expressions, we refer the reader to App.~\ref{app:Nmgauge} (see also~\cite{Fidler:2017pnb}). 

Once we have specified the Nm gauge, we can construct the coordinate system that is implicitly assumed in a Newtonian simulation and obtain a fully relativistic solution from the Newtonian one.

In the limit of a pure cold dark matter cosmology, a particularly simple Nm gauge can be obtained. In this case, it can be shown that there exists a Nm gauge with $\HT = 3 \zeta$ at all times, where $\zeta$ is the spatial curvature perturbation. Such a gauge achieves formal agreement between the relativistic and Newtonian equations of motion, and was named {\it $N$-body gauge} in the literature; see Ref.\,\cite{Fidler:2015npa} but note that a different time slicing has been used. 
Instead, the gauge that we are using shares the temporal gauge condition with the {\it Poisson gauge}, and this is why that gauge has been called {\it N-boisson gauge} in~\cite{Fidler:2018bkg}.

The N-boisson gauge describes the relativistic coordinate system that can be implicitly assumed in all standard Newtonian cold dark matter N-body simulations. Throughout this paper we will use it as a reference to make our massive neutrino simulations comparable with ordinary matter simulations. 
In the presence of massive neutrinos, we require the Nm gauge to coincide with the N-boisson gauge at initial time. The Nm gauge condition is then solved starting from $\HT\rvert_{t_{\rm ini}} = 3 \zeta$, and neutrino perturbations gradually push the Nm gauge away from N-boisson gauge as time progresses. Therefore, the N-boisson gauge will often serve as a point of reference in order to quantify the effect of additional species, such as massive neutrinos and photons. 
We note that instead of fixing the space-time to N-boisson gauge at initial time, other choices for $\fett{L} \rvert_{t_{\rm ini}}$ and $\partial_t \fett{L}\rvert_{t_{\rm ini}}$ are possible (or, equivalently for $\HT \rvert_{t_{\rm ini}}$ and $\dot{H}_{\rm T} \rvert_{t_{\rm ini}}$); see 
section~\ref{sec:other_stuff} and the accompanied appendix~\ref{app:backwards}.

The reason that the Nm method is very efficient is that the coordinate shift $\fett{L}$ turns out to be independent of the non-linear matter dynamics in the simulation. Furthermore, for the range of neutrino masses that is currently favoured by observations ($\sum m_\nu < 0.12$\,eV~\cite{Aghanim:2018eyx}), neutrino perturbations remain close to linear at all times. We can therefore find an accurate solution for the Nm gauge condition using linear Boltzmann solvers such as {\sc class} \cite{class2} or {\sc camb} \cite{Lewis:1999bs}. 
The presented Nm approach should especially be contrasted with grid-based  approaches such as the \COSIRA{} method
of~\cite{Tram:2018znz} (see also~\cite{Brandbyge:2016raj}). There, equation~\eqref{relatEOM} is solved by determining the non-Newtonian term $\delta \fett{F}$, which contains the effect from neutrinos, from a modified version of
the linear Boltzmann code \CLASS{}. A realisation of~$\delta \fett{F}$ is sampled on the lattice of the simulation at each time step, and thereby included in the Poisson solver. While the \COSIRA{} method in principle allows for including the non-linear feedback of neutrino perturbations on the dynamics of particles in the simulation, we will show that the impact of neutrino perturbations is sufficiently small on the non-linear scales, such that the Nm approach is valid even without a modification of the Poisson equation at the run-time of the simulation (see figure \ref{fig:H_T_forward}). Nevertheless, the presence of massive neutrinos changes the expansion rate in the simulation, which must be taken into account with both methods. This change in the expansion rate leads to a non-perturbative feedback to the dynamics in the simulation, that affects all scales. 

Thus, by applying the Nm framework, we can merge linear perturbation theory for neutrinos and general relativity with the non-linear N-body simulation. In this way we include the non-perturbative feedback of massive neutrinos on structure formation, while we only neglect the non-linearities in the neutrino distribution which would become important on small scales. 
However, in comparison to matter, the neutrino power is suppressed by several orders of magnitude on small scales due to free-streaming, which delays the neutrino clustering on these scales. For this reason, matter completely dominates over the neutrinos on small scales, even when including the small non-linear enhancement of those \cite{Adamek:2017uiq}. As we will show, the presented approach remains valid up to high accuracy for combined neutrino masses up to at least $\mathcal{O}(0.5 \, \rm eV)$, and thus an explicit representation of the neutrino phase-space in the simulation can be avoided. Since in the Nm approach the corrections coming from general relativity and massive neutrino perturbations are incorporated in the coordinate shift $\fett{L}$, a Newtonian simulation can be turned into a weak-field relativistic simulation with massive neutrinos as a kind of post-processing; for details see the following section.

\section{Implementation}
\label{sec:Method}

Here we outline the detailed steps to employ the Newtonian motion (Nm) gauge approach in combination with an ordinary Newtonian N-body simulation --- the non-expert reader may skip this section and go directly to the results section~\ref{sec:results}. We will focus on the so-called forwards method~\cite{Fidler:2018stc}; see section~\ref{sec:other_stuff} for an alternative implementation.

\begin{figure}
\centering
  \includegraphics[width=0.95\linewidth]{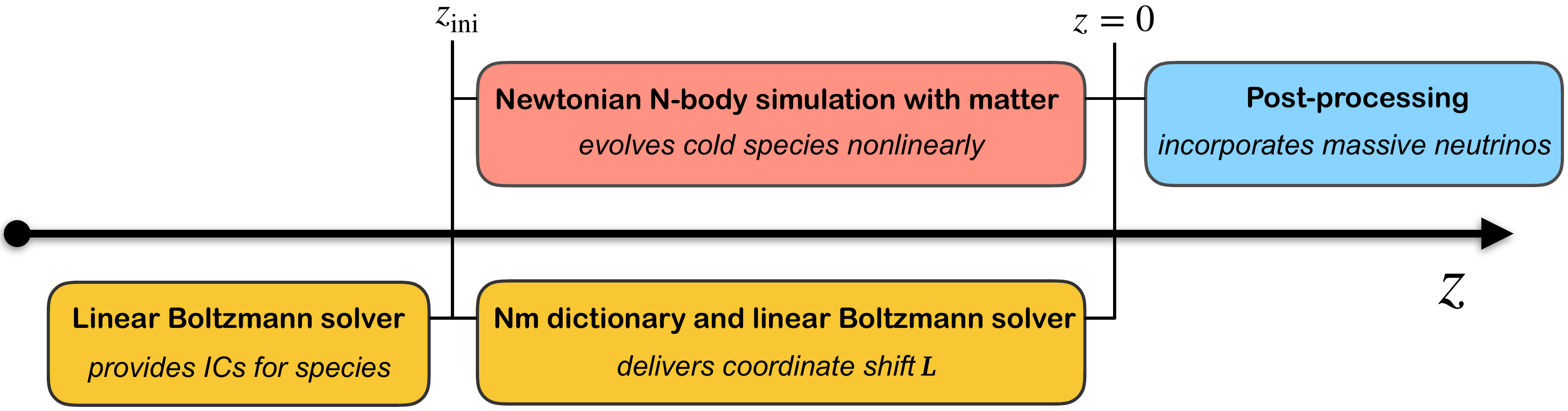} 
  \caption{Roadmap for performing N-body simulations using the Nm approach. After evolving the linear fluid variables  up to $z_{\rm ini}$  (we set $z_{\rm ini} = 100$) with a Boltzmann code, we generate particle displacements and velocities according to Eqs.\,\eqref{eqn:v_n_ini}. 
This provides initial conditions for the N-body simulation which is used to evolve the cold species nonlinearly from $z_{\rm ini}$ until $z=0$. Whilst being completely decoupled from the N-body simulation, we determine the respective Nm quantities, in particular the coordinate shift $\fett{L}$. Once the simulation is completed, we transform its output to any gauge for further analysis (e.g., N-boisson gauge which is close to a Newtonian setup), using $\fett{L}$.}
  \label{fig:nm_recipe}
\end{figure}

Our method, which is schematically summarised in figure~\ref{fig:nm_recipe}, is as follows. Using {\sc class} we compute the linear transfer functions in Poisson gauge for all species that are present in the desired cosmology up to a redshift of $z_{\rm ini} = 100$. Note specifically, that in the present approach we do {\it not} use the $z=0$ output of the Boltzmann code and scale its amplitudes back to $z_{\rm ini}$ using the scale independent linear growth of matter, $D_{+}$. Instead we take directly the Boltzmann output at $z_{\rm ini}$ (see section~\ref{sec:other_stuff} where we use another procedure). 
This serves as the input to generate the initial conditions for the cold species according to Eq.\,(\ref{eqn:v_n_ini}). Subsequently, the cold species are then non-linearly evolved using a Newtonian simulation (in this work, either \emph{gevolution} \cite{Adamek:2015eda, Adamek:2016zes} or {\sc gadget-2} \cite{Springel:2005mi}). Finally we apply the above outlined post-processing to transform the output from the Nm gauge to the N-boisson gauge, thereby making the results comparable to that of ordinary N-body simulations.

The approach for incorporating massive neutrinos relies on only three minor modifications to a standard Newtonian N-body analysis, namely (1) preparing the initial conditions for the simulation in the Nm gauge~(section~\ref{ICsNM}); (2) adding the background effect of massive neutrinos into the Newtonian code~(section~\ref{Friedmann}); and (3) extracting the coordinate shift $\fett{L}$ (or the off-diagonal spatial metric perturbation $\HT$) from a linear Boltzmann code and use it to  post-process the output of the simulation~(section~\ref{postpro}). 

\subsection{Initial conditions in the Nm approach}\label{ICsNM}

To generate the initial conditions for the matter component, we use as input the relativistic density and velocity in Poisson (P) gauge as computed in linear Boltzmann codes. 
(We use the notation that simulation quantities carry a superscript 'N', while relativistic perturbations carry their gauge as a superscript.) 
From the relativistic density $\delta_{\rm m}^{\rm P}$ and velocity divergence $\theta_{\rm m}^{\rm P}$ 
we determine the respective Newtonian counterparts that are needed to initialise the matter simulation, 
\begin{equation}
\begin{aligned}
  \delta^{\rm N} &= \delta_{\rm m}^{\mathrm{P}} - H_{\rm T} + 3\Phi \,,   \\
  \theta^{\rm N} &= \theta_{\rm m}^{\rm P}  +  \dot H_{\rm T}  \,, \label{eqn:v_n_ini}
\end{aligned}
\end{equation}
where $\Phi$ is the Bardeen potential
(cf.\ \cite{Ma:1995ey}) and a dot denotes a derivative
w.r.t.\ conformal time.  These relations are derived from the analysis
as outlined in \cite{Fidler:2016tir}, and, supplemented with appropriate boundary conditions, serve as the input for determining the coordinate shift $\fett{L}$. For example, when initialised in the N-boisson gauge, we choose $\fett{L}\rvert_{t_{\rm ini}}=0 =\partial_t \fett{L}\rvert_{t_{\rm ini}}$.
On the level of the metric potentials, these initial conditions translate into $\HT \rvert_{t_{\rm ini}} = 3 \zeta$ and $\dot{H}_{\rm T} \rvert_{t_{\rm ini}} = 3 \dot{\zeta}$.
We note that, as opposed to
standard initial conditions for simulations, in the Nm approach initial particle positions and velocities are not initialised from a shared (density) growing mode; instead they are initialised seperately, using first-order space-like and velocity-like displacements, both drawn from the same random realisation.

\subsection{Friedmann equation}\label{Friedmann}

In addition to fixing the initial conditions for the simulation, particles in Newtonian simulations experience the Hubble drag and thus, the matter trajectories are directly affected by changes in the Friedmann equation.
Compared to a pure cold dark matter simulation, the presence of massive neutrinos does alter the Hubble rate, since neutrinos transit from relativistic to non-relativistic behaviour at late times. This transition induces a fully non-perturbative effect on the matter trajectories and thus must be taken into account.
Therefore, we update the Friedmann equation in the N-body simulation with 
\be
  [H(a)/H_0]^2 = \Omega_{\rm m0}\, a^{-3} +  \Omega_{\nu}(a)  + \Omega_{\gamma0} \, a^{-4} + \Omega_{\Lambda0} \,,
\ee
where the index '0' denotes quantities evaluated at the present time, and 
$({\rm m}, \nu, \gamma, \Lambda) =$ (cold dark matter + baryons, massive neutrinos, radiation, cosmological constant). The computation of $\Omega_{\nu}(a)$ requires the integration of the neutrino phase-space at the background level. 
To ensure sufficiently high precision, which is especially relevant for avenues beyond standard $\Lambda$CDM, it is advisable to use well established tools such as {\sc class} or {\sc camb}.
We note that the evolution of $\Omega_{\nu}(a)$ depends sensitively on the individual masses of the neutrinos~(see table~\ref{tab:parameters} for our choice of cosmological parameters).

\renewcommand{\arraystretch}{1.2}
\setlength{\tabcolsep}{16pt}

\begin{table}[t]
 \centering
  \begin{tabular}{ c | c c c c c}
    $\sum m_\nu$ & $m_{\nu 1}$     & $m_{\nu 2}$      & $m_{\nu 3}$     &  $N_{\rm ur}$  &   $\omega_{\rm cdm}$  \\
   \hline
              0 &    0            &   0            &    0           &   3.046      &  0.120380 \\ 
       0.1\,eV  &    0.0225\,eV    &  0.0225\,eV    &   0.0550\,eV  &   0.00641    &  0.119306\\
       0.2\,eV  &    0.0600\,eV   &   0.0600\,eV   &    0.0800\,eV  &   0.00641    &  0.118233 \\
       0.3\,eV  &    0.1000\,eV      &   0.1000\,eV       &    0.1000\,eV      &   0.00641    &  0.117159 
  \end{tabular}
\caption{Used cosmological parameters for our simulations, which were initialised at $z_{\rm ini}=100$. For all runs we set $A_s = 2.215 \cdot 10^{-9}$, $n_s = 0.9619$, $T_{\rm cmb} = 2.7255$\,K, and $h = 0.67556$. To have the same matter content in the reference $\mathrm{\Lambda}$CDM and massive neutrino cosmology at final time, the CDM mass density of the massive neutrino cosmologies is decreased by $\omega_\nu = \sum m_\nu/(93.14\,\mathrm{eV})$. }
\label{tab:parameters}
\end{table}

\renewcommand{\arraystretch}{1}
\setlength{\tabcolsep}{6pt}

\subsection{Post-processing}\label{postpro}

After the simulation has been completed,
we post-process the output using a coordinate transformation to convert the N-body simulation snapshot to any gauge that is required for further analysis.
For example, to transform the particle positions to the gauge that most closely resembles a Newtonian setup, i.e., the N-boisson (Nb) gauge, we apply

\begin{align}\nonumber
    \fett{x}^{\mathrm{Nb}} &= \fett{x}^{\mathrm{Nm}} + \fett{L}^{\mathrm{Nm}\rightarrow \mathrm{Nb}} \,, \\
    \nab \cdot \fett{L}^{\mathrm{Nm} \rightarrow \mathrm{Nb}} &= H_{\rm T} - 3 \zeta =: \xi^{\rm post} \label{post-trafo}
\end{align}
to all final particle positions, where $\fett{L}^{\mathrm{Nm}\rightarrow \mathrm{Nb}}$ is the relevant coordinate shift that depends on the metric perturbation $\HT$ and on the comoving curvature perturbation $\zeta$ which we both determine from the Boltzmann code {\sc class}. 

Details on the numerical computation of the metric perturbation $\HT(k)$ 
are given in appendix~\ref{app:Nmgauge} (see in particular equation~\ref{gaugedefinition}).
The output of the Boltzmann code, i.e., the Fourier transform of $\xi^{\rm post} = H_{\rm T} - 3 \zeta$, only depends on the modulus of the Fourier mode~$\fett{k}$. Because $\xi^{\rm post}$ is the divergence of the coordinate shift, the Fourier space output can be used to generate a real-space realisation of $\fett{L}^{\mathrm{Nm} \rightarrow \mathrm{Nb}}$ from the same white-noise field $\mathcal{R}(\boldsymbol{k})$ that is also used for the initial conditions (see e.g.\ appendix A in~\cite{Dakin_2019}). The so-obtained coordinate shift in real space $\fett{L}^{\mathrm{Nm} \rightarrow \mathrm{Nb}}(\fett{x})$ is then used to update the particle positions in the simulation\footnote{For the presented results, we used the Lagrangian positions (rather than the Eulerian coordinates). However, because $\fett{L}^{\mathrm{Nm} \rightarrow \mathrm{Nb}}(\fett{x})$ has no support on small scales (i.e., $\fett{L}^{\mathrm{Nm} \rightarrow \mathrm{Nb}}(\fett{x})$ = $\fett{L}^{\mathrm{Nm} \rightarrow \mathrm{Nb}}(\fett{q})$), this distinction is not important here.}. 

The computational complexity of this post-processing is trivial: it is identical to the generation of initial conditions in the Zel'dovich approximation and consists of a 3-dimensional Fourier transformation of the linear perturbation $\xi^{\rm post}(k)$ multiplied by the white noise $\mathcal{R}(\boldsymbol{k})$.

\begin{figure}
\centering
  \includegraphics[width=0.8\linewidth]{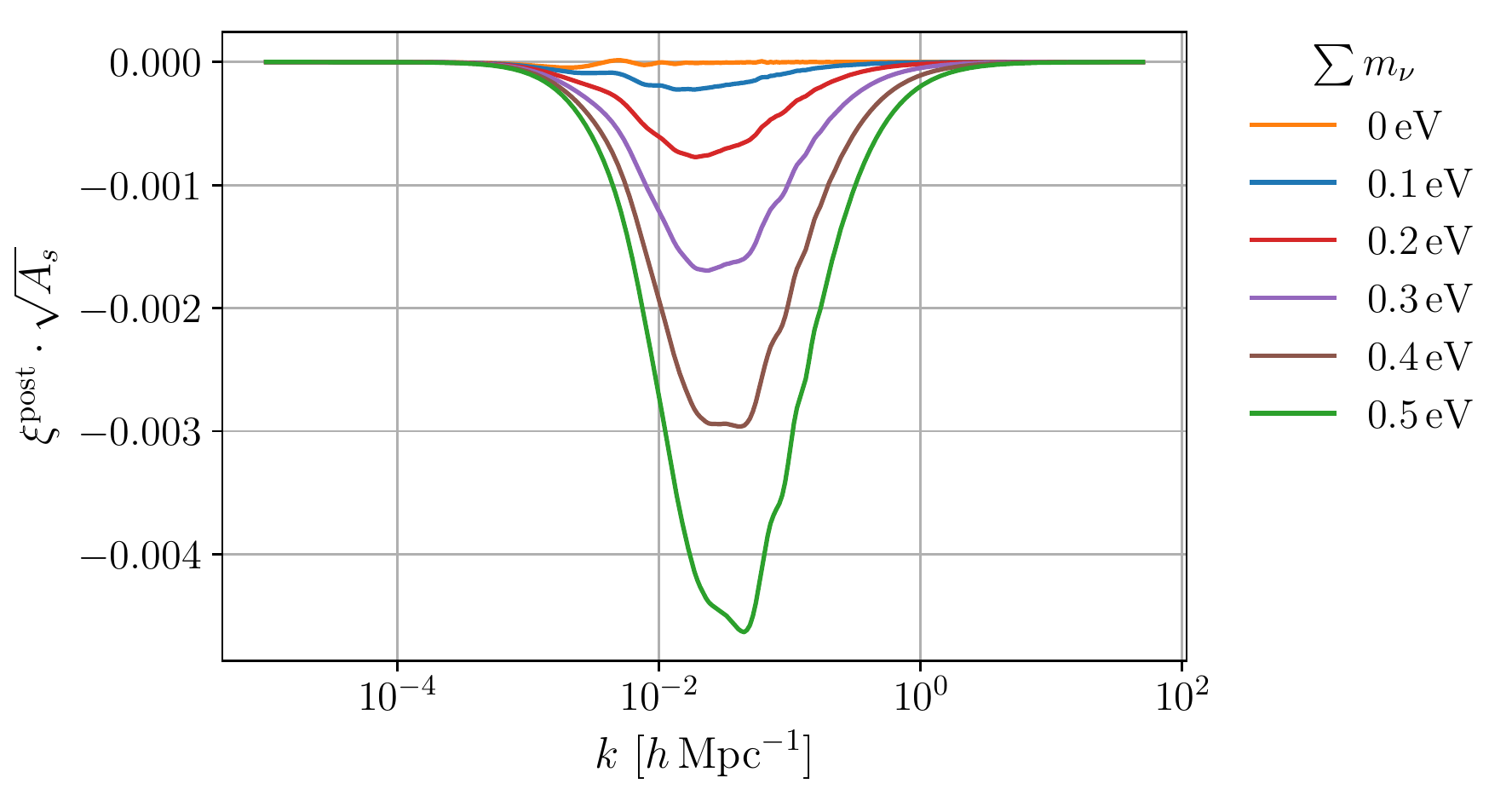}
  \caption{Shown is the integrated effect of neutrinos and photons in comparison to a cosmology without any external species as a function of scale $k$ at $z=0$, quantified by the post-processing displacement $\xi^{\rm post}$ that needs to be applied at the end of the simulation. 
In order to assess the underlying smallness assumption of our weak-field approach, we have multiplied $\xi^{\rm post}$ by the scalar normalisation $\sqrt{A_s}$, with $A_s = 2.215 \cdot 10^{-9}$.
The orange line is for a massless neutrino cosmology and describes the impact of relativistic species (photons and massless neutrinos) (cf.\,\cite{Adamek:2017grt}). Massive neutrinos significantly enhance $\xi^{\rm post}$ (coloured lines) with a peak on scales around $k \approx 10^{-2} h \, {\rm{Mpc}}^{-1}$ while their impact is suppressed on small scales due to their free-streaming property. As it is evident from the absolute smallness of $\xi^{\rm post}$, the maximum of the post-processing is smaller than $\sim0.5\%$ for a combined mass of $\sum m_\nu = 0.5$\,eV and is therefore compatible with the weak-field approach.}

  \label{fig:H_T_forward}
\end{figure}

Crucially, our Nm approach remains accurate as long as the metric perturbation $\HT$ remains sufficiently small so that it does not violate the underlying weak-field assumptions. We find this to be the case for $\Sigma \,m_\nu \lesssim 0.5$\,eV; see the following results section (see Fig.\,\ref{fig:H_T_forward}).

\section{Results} \label{sec:results}

Here we present the results obtained with the Nm method applied to a standard Newtonian N-body simulation. As discussed in the previous section, the method relies on minimal modifications in the initial conditions of standard matter simulations (provided at $z_{\rm ini}=100$), as well as a change in the Hubble rate thereof. Once these simulations are completed, a computationally very cheap post-processing is required to include the impact of neutrinos.

All simulations were performed using the cosmological parameters shown in table~\ref{tab:parameters}. To keep the content of non-relativistic matter at final time fixed, we decrease the dark matter density by $\omega_{\rm cdm}^{\mathrm{massive}} = \omega_{\rm cdm}^{\mathrm{ref}} - \sum m_\nu/(93.14\,{\rm eV})$. 
In order to compute the relative power spectra with the particle mesh gravity solver of the code  \emph{gevolution} we combine the results of runs with multiple different box-sizes. With this strategy, we can test convergence on non-linear scales (by comparing the results from runs with different box-sizes), but we can also check the accuracy of the gauge-transformation applied to the particle snapshot on large scales by comparison to the linear expectation --- without limitations due to cosmic variance. 
For each neutrino mass sum, we employ 7 different box sizes: For $L_{\mathrm{box}} = \{16, 8, 4, 2, 1, 0.5 \}\, \mathrm{Gpc}/h$, we use $N_p = 2048^3$ matter particles. The last run for $L_{\mathrm{box}} = 64\, \mathrm{Gpc}/h$ only uses $1024^3$ particles and is just intended to prove the agreement of our method with linear theory. To combine the results from the different box-sizes, the individual power spectra are cut at $5 \, k_{\rm min}$ in order to reduce cosmic variance ($k_{\rm min} = 2 \pi/L_{\rm box}$) and $0.2 \, k_{\rm Ny}$, to avoid incorrect results close to the Nyquist frequency $k_{\rm Ny} = \pi N_p/L_{\rm box}$. It is important to note that some resolution effects close to the Nyquist frequency cancel in the ratio of power spectra (cf.\,\cite{Adamek:2017uiq}). 
To merge the power spectra from these runs, we average the results in overlapping $k$-regimes. However, in the regions of overlap, the power spectra agree to permille accuracy (proving the convergence of the relative power spectra) and it is therefore in general possible to obtain similar results with fewer different box-sizes, especially if methods with more small-scale resolution are employed. In general, it is also possible to employ smaller boxes than in this work and to resolve smaller scales. However, because the impact of neutrino perturbations is suppressed on small scales (see e.g.\ \cite{Hannestad:2020rzl}), the only neutrino effect on these scales originates from the change in the Hubble friction and neutrino simulations for the total-matter distribution become trivial.

The post-processing that delivers the coordinate shift $\fett{L}$ is fully specified by the metric potential $\HT$ (and the gauge invariant curvature perturbation $\zeta$). As mentioned in section~\ref{sec:recap}, in the absence of neutrino and photons, the so-called N-boisson gauge satisfying $\HT = 3 \zeta$ is an exact solution of the Nm gauge condition~\cite{Fidler:2017ebh}. In a more realistic cosmology, corrections stemming from neutrino and photon interactions drive $\HT$ away from its N-boisson gauge value, and the Nm metric potentials acquire a non-trivial time evolution. The difference between $\HT$ and $3\zeta$ is exactly $\xi^{\rm post}$ as defined in Eq.\,\eqref{post-trafo}, which we show in figure~\ref{fig:H_T_forward} for a massless and massive neutrino cosmology at final time $z = 0$. The figure clearly indicates that massive neutrinos have a significantly larger impact on $\xi^{\rm post}$  than photons and peak on intermediate scales, where the neutrino density transfer function takes its maximum amplitude. On small scales, $\xi^{\rm post}$ is suppressed due to the neutrino free streaming.

As mentioned earlier, it is crucial to note that the Nm approach is accurate only as long as the metric perturbations remain small in the weak-field sense. Since the post processing $\xi^{\rm post}$ is directly related to the metric perturbations, we should therefore demand that it remains small compared to the square root of the scalar amplitude $A_s$ at all times. For larger neutrino masses the absorbed corrections become increasingly important and we therefore expect a growing $\xi^{\rm post}$ that may eventually conflict with our assumptions. We have however explicitly verified, that the underlying weak-field limit is accurate until at least $\sum m_\nu \lesssim 0.5$\,eV, see figure~\ref{fig:H_T_forward}.

\begin{figure}
\centering
  \includegraphics[width=0.85\linewidth]{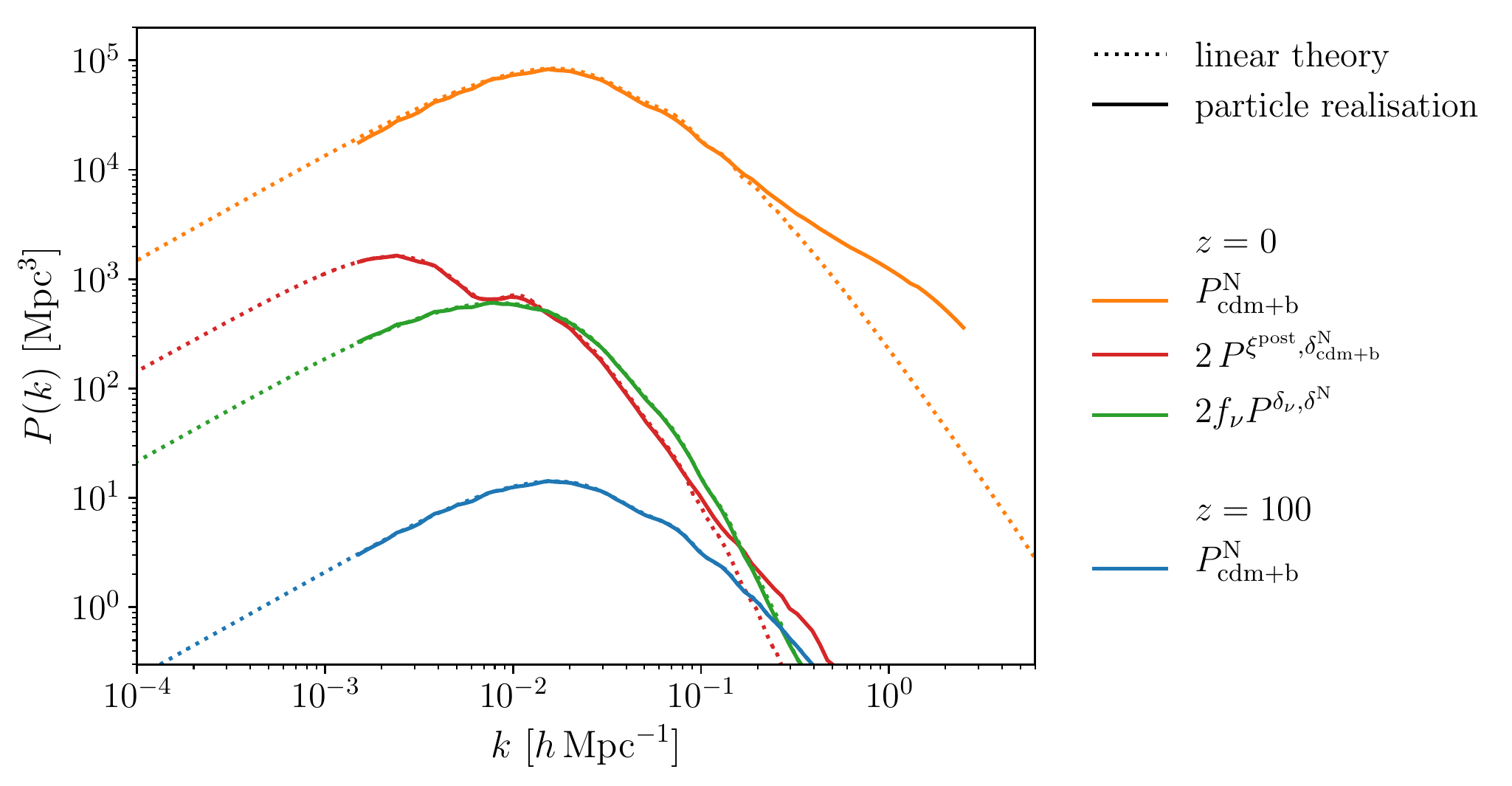}
 \caption{Shown are the initial and final matter power spectra from our Newtonian simulation in solid lines versus the linear theory in dotted lines for $\sum m_\nu = 0.1$\,eV. The blue line is the initial power spectrum, which closely resembles that of the synchronous gauge at $z_{\rm ini}$. The orange curve shows the simulation density as it is after evolving to $z = 0$ under Newtonian gravity. In our analysis we still need to post-process the densities by displacing particles to incorporate the impact of the massive neutrinos, adding the red power spectrum. (The factor 2 is due to the symmetry in the cross-spectrum). This correction is particularly relevant for scales larger than $k = 10^{-2} h  {\rm Mpc^{-1}}$. Finally, the linear neutrino densities are incorporated in the analysis, leading to a cross-correlation with the nonlinear matter density (green line), where $f_\nu = \Omega_{\nu0}/\Omega_{\rm m0}$.} 	    
 \label{fig:p_k_forward}
\end{figure}

\begin{figure}
\centering
  \includegraphics[width=0.86\linewidth]{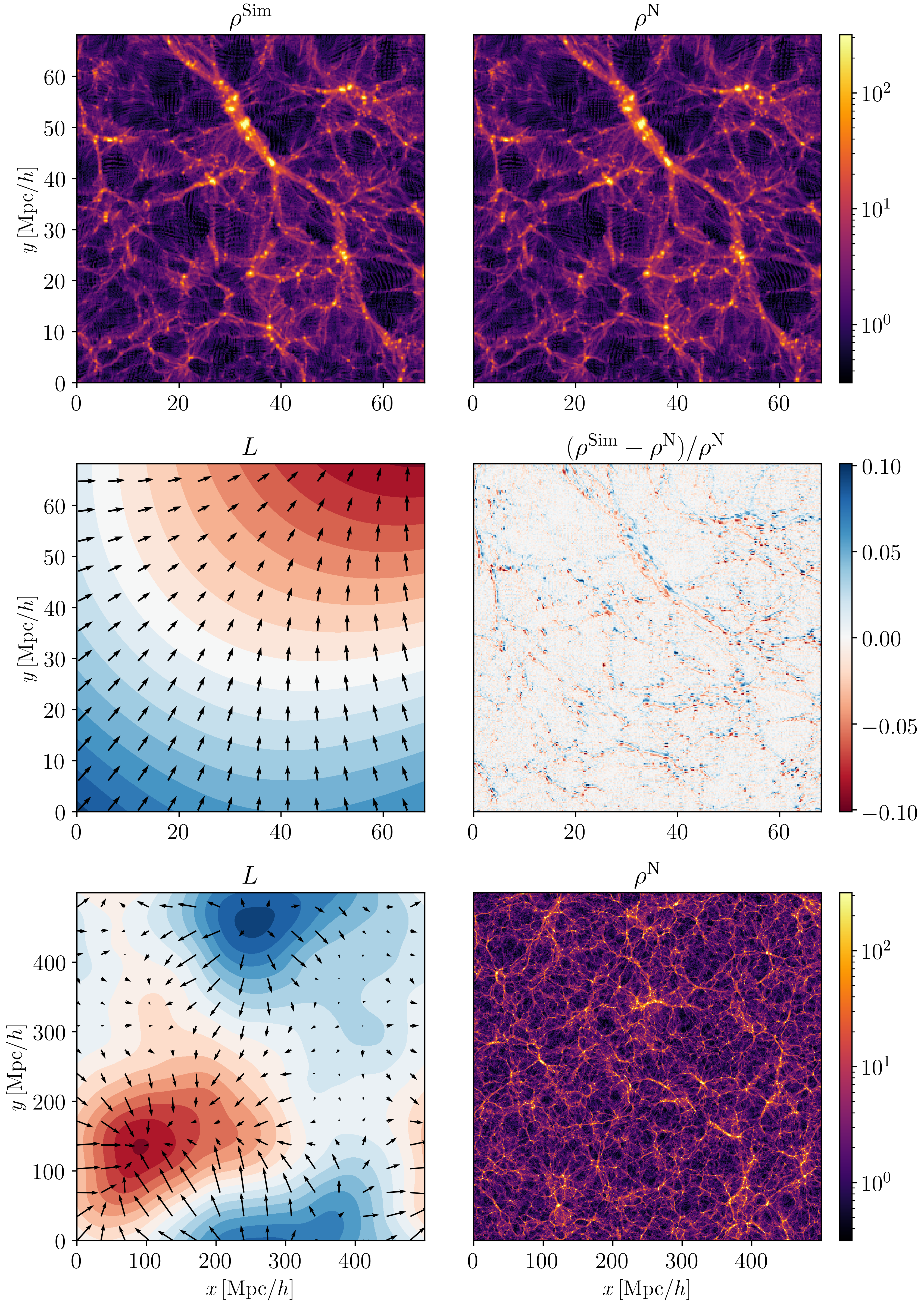}

  \caption{Comparison between simulation densities before and after the post-processing in a small simulation volume with box length 70\,Mpc$/h$ (first row) for $\sum m_\nu = 0.1$\,eV. On small scales, the impact of the post-processing is a homogeneous shift $\fett{L}$ (second row), which results in a dipole-like feature in the relative densities. On large scales (box length 500\,Mpc$/h$; bottom-left panel), the coordinate shift  $\fett{L}$ is inhomogeneous, leading to significant corrections for the power spectrum on large scales~(cf.\,Fig.\,\ref{fig:p_k_forward}).}
  \label{fig:displacementfield} 
  
\end{figure}

In figure~\ref{fig:p_k_forward} we present results for the powerspectrum as measured from our simulations for $\sum m_\nu = 0.1$\,eV. The blue line is the initial matter power spectrum according to equation~\ref{eqn:v_n_ini}, which is almost identical to the synchronous gauge power spectrum that is commonly used to initialise ordinary N-body simulations.\footnote{Note again that we do not perform any backscaling here. The simulations are used to evolve the cold species until the final time ($z=0$), yielding the orange curve. Due to the presence of massive neutrinos in our cosmology, the coordinate shift $\fett{L}$ and its divergence $\nabla \cdot \fett{L} = \xi^{\rm post}$ have acquired non-negligible contributions;
the latter leads to an effective change of the simulation density, i.e., $\delta \to \delta^{\rm N} + \xi^{\rm post}$.}  On the level of power spectra, this post-processing mainly enters via the cross-spectrum between $\delta^{\rm N}$ and $\xi^{\rm post}$,
namely $\langle (\delta^{\rm N}+ \xi^{\rm post})^2 \rangle \supset  2\langle \delta^{\rm N} \xi^{\rm post} \rangle \sim 2 P^{\xi^{\rm post},\delta_N}_{\rm cdm+b}$  
which is shown in red in figure~\ref{fig:p_k_forward}, thereby adding effectively corrections at the percent-level on large and intermediate scales (while $\langle \xi^{\rm post} \xi^{\rm post} \rangle$ is suppressed at all relevant scales).
On small scales, by contrast, those corrections are negligible. As a last step of the post-processing we finally add a linear realisation of the present day neutrino power spectrum using the same random seed as for the initial conditions in order to obtain the total matter power spectrum. Again the cross-term is dominant, shown in green.

\begin{figure}[t]
\centering
  \includegraphics[width=0.85\linewidth]{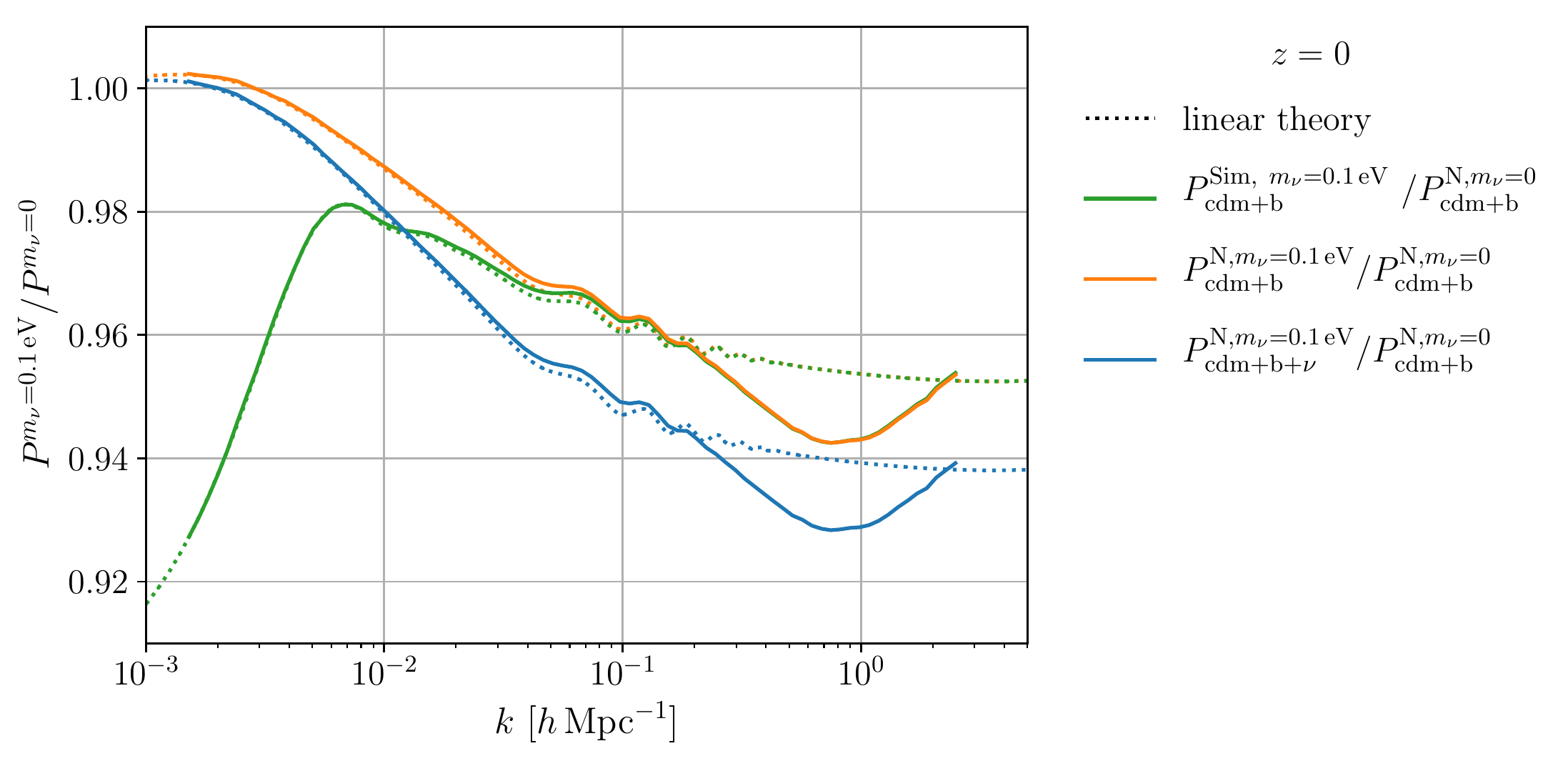}
  \caption{Plot of the relative difference in power between a massive and massless neutrino cosmology for $\sum m_\nu =0.1$\,eV. The green line is what is originally measured from the CDM Newtonian simulation. We then apply our post-processing to find the orange line for the power spectrum of cold dark matter and baryons. In order to compare with the total matter power spectrum we finally add the linear neutrino perturbations to obtain the blue line. We find the expected neutrino suppression with the typical spoon shaped contour.}
  \label{fig:suppression_forward}
\end{figure}

The impact of the post-processing on the level of real-space densities is shown in figure~\ref{fig:displacementfield}. The top panel shows the comparison of the density distribution before (left panel) and after  (right panel) we applied the coordinate shift $\fett{L}$, the latter being explicitly shown in the central-left panel. Note again, that the simulation density $\rho^{\rm N}$ as shown in the upper left plot does not include the effect of neutrino perturbations yet, while the neutrino background density is already included in the Friedman equation; in our approach the direct impact of neutrino perturbations comes in only through the post-processing. As a consequence of neutrino free-streaming, the post-processing is almost homogeneous on small scales ($L_{\mathrm{box}} = 70 \,\mathrm{Mpc}/h$) and leads to a dipole-like shift in the relative densities (central-right panel). However, the impact of the post-processing is barely visible by eye. On larger scales ($L_{\mathrm{box}} = 500\,\mathrm{Mpc}/h$), it becomes evident that the impact of $\fett{L}$ is inhomogenous. Therefore, our postprocessing operation shifts the position of small-scale structures while keeping their internal structure basically unchanged.

In Fig.\,\ref{fig:suppression_forward} we show the ratio between power spectra at $z=0$ with massless neutrinos versus massive neutrinos~($\sum m_\nu =0.1$\,eV). The green curve corresponds to the unprocessed result of the N-body simulation, where $\xi^{\rm post}$ is not yet applied, while the orange curve includes it. In both cases we recover the result of linear theory on large scales. On small scales, where the correction from $\xi^{\rm post}$ becomes negligible, the post-processing becomes redundant.
This directly shows that for light neutrinos the spoon shaped suppression is an effect that is essentially due to the expansion history of the Universe, but not to the neutrino density perturbations: massive neutrinos delay the time of matter-radiation equality if the amount of non-relativistic matter at $z=0$ is held constant \cite{Lesgourgues:2012uu}. Since dark matter can only begin to collapse efficiently during matter domination, the growth of structures in a massive neutrino cosmology is delayed already at the linear level. As a consequence, in the massive neutrino cosmology, modes also enter the non-linear regime later which leads to an additional suppression on scales smaller than $k \gtrsim 10^{-1} h / {\rm Mpc}$. Finally,  the blue curve in Fig.\,\ref{fig:suppression_forward} includes the linear densities of massive neutrinos, computed by \CLASS{}, yielding the final result in the approach, the total matter power spectrum.

\begin{figure}[t]
\centering
  \includegraphics[width=0.92\linewidth]{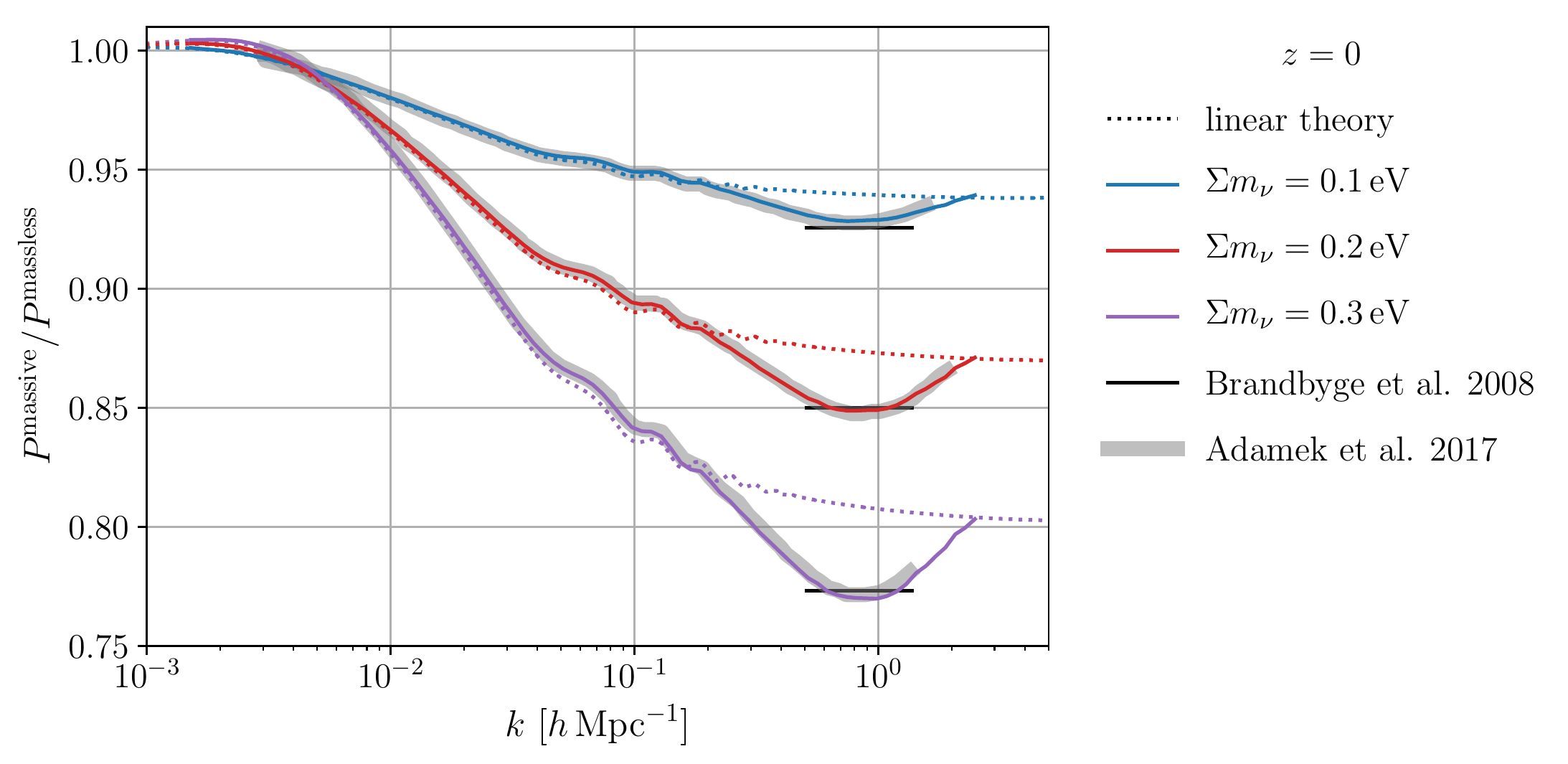}
 \caption{Suppression of the matter power spectrum due to massive neutrinos. We compare our simulation method (thin colored solid lines) against the simulations as carried out in~\cite{Adamek:2017uiq} (thick lines in grey). Note that we use  $2048^3$ particles for five different box sizes, while Ref.\,\cite{Adamek:2017uiq} uses $4096^3$ matter particles for one box only. 
For reference, we also show the fitting function $\Delta P/P|_{\max } = -9.8 f_\nu$ for the power suppression, as derived in~\cite{Brandbyge:2008rv} (black horizontal lines).}
  \label{fig:suppression_forward-comparison}
\end{figure}

In Fig.\,\ref{fig:suppression_forward-comparison} we compare our results for the power spectrum suppression  at $z=0$ for various values of $\sum m_\nu$ against current state-of-the-art simulations, as carried out in Ref.\,\cite{Adamek:2017uiq}. These reference simulations are relativistic in the weak-field sense, and evolve massive neutrinos actively by sampling neutrino particles. As it is evident, our results agree with this method to permille accuracy for all neutrino masses that have been considered in Ref.\,\cite{Adamek:2017uiq}. 
Note again that here we only use $2048^3$ for various different simulation box sizes, while Ref.\,\cite{Adamek:2017uiq} uses $4096^3$ dark matter and $5540^3$ neutrino particles for one simulation box only. As already discussed, by cutting the individual power spectra well before the Nyqist frequency and gluing them together, we establish numerically converged results to permille accuracy.

The position of the dip is also in agreement with the findings of \cite{Brandbyge:2008rv}, where the fitting function 
$\Delta P/P|_{\max } \approx -9.8 \,f_\nu$ was derived, with $f_\nu :=  \Omega_{\nu0}/\Omega_{\rm m0}$. However, this fitting function is empirical and for example insensitive to the mass splitting, which can have an impact on the power spectrum of several permille as shown in Ref.\,\cite{Adamek:2017uiq}.

In conclusion, Fig.\,\ref{fig:suppression_forward-comparison} demonstrates that the Nm method for massive neutrinos produces accurate results for masses up to at least $\sum m_\nu = 0.3$\,eV, which is the highest mass that was considered in \cite{Adamek:2017uiq}. 
We note that although such large masses are already disfavoured by observations, a sufficiently large mass regime for the neutrinos should be allowed for parameter extraction (e.g.\ for MCMC sampling).\footnote{We thank Yvonne Y.Y.\ Wong for pointing this out.}

\section{Further results} 
\label{sec:other_stuff}

\subsection{Alternative implementation}\label{alternativeimpl}

As already discussed in section~\ref{sec:recap}, there is not one unique Nm gauge, but rather a class of gauges defined by the initial conditions for the Nm gauge equation. In the previous chapters, 
we have chosen an Nm gauge by requiring that it coincides with the N-boisson gauge at initial time. This choice is motivated by the fact, that in this case $\xi^{\rm post}$ directly singles out the effect of neutrino and photon perturbations along the CDM trajectories. We call such a choice of initial conditions the \textit{forwards} approach. However, it is possible to design other Nm gauges that produce physically equivalent results, but do not require a post-processing. In this \textit{backwards} method, the freedom of choice for the initial conditions is used in such a way that $\xi^{\rm post}$ is minimised. Therefore, the output of the simulation at final time is directly comparable to that of an ordinary N-body simulation without the need of a post-processing, however at the price of starting from a special set of initial conditions. Hence, the final time post-processing can be avoided in exchange for a more sophisticated set of initial conditions, that will non-linearly affect the following Newtonian evolution. These initial conditions can be found iteratively with the procedure explained in appendix~\ref{app:backwards}. In the case of a massless neutrino cosmology this is equivalent to the commonly employed backscaling approach, while it does not have such a simple interpretation for massive neutrinos; see~\cite{Fidler:2018stc} for details.

\begin{figure}
\centering
  \includegraphics[width=0.9\linewidth]{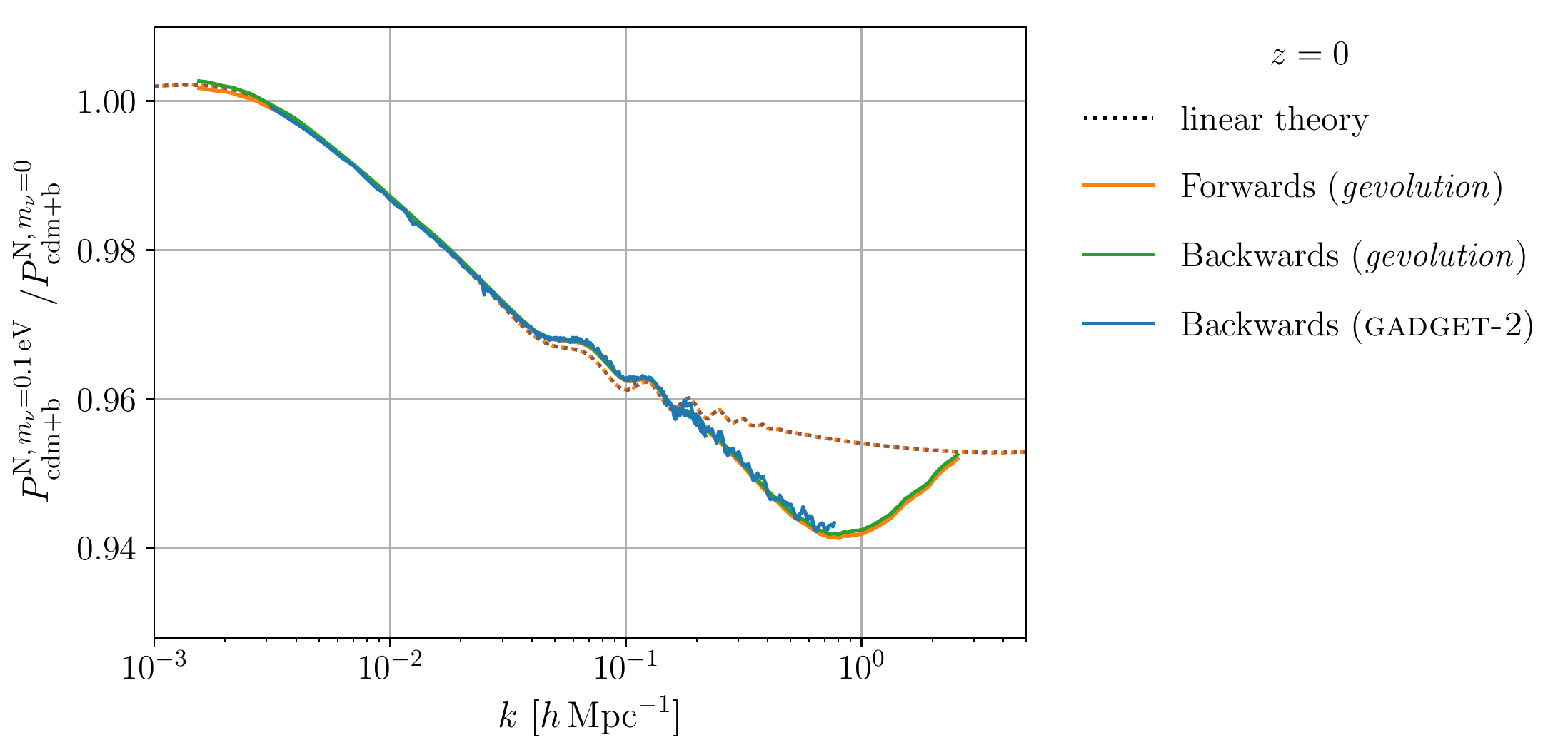}
  \caption{Shown are the relative power spectra using the backwards method (section~\ref{sec:other_stuff}) within \emph{gevolution} (green line, $N_p = 2048^3$) and {\sc gadget-2} (blue line,  $N_p = 512^3$). By construction, the backwards method does not require any post-processing at the final time, as opposed to the forward approach (orange line) which is employed in the previous sections. }
  \label{fig:back_spoon}
\end{figure}

In figure~\ref{fig:back_spoon}, we demonstrate that the forwards method after the post-processing produces the same results in N-boisson gauge as the backwards method, that does not require a post-processing. 
This illustrates, that physical results are independent from the chosen Nm gauge. We also implemented the backwards method in \textsc{gadget-2}, where we only use $N_p = 512^3$ particles in comparison to $N_p = 2048^3$ with \emph{gevolution}. However, due to the TreePM algorithm employed in \textsc{gadget-2}, the small-scale resolution is much higher than in a pure particle mesh code. Therefore, the results obtained with both codes coincide almost perfectly even with the smaller particle number in \textsc{gadget-2}.

\subsection{Lightcone analysis}
\label{sec:lightcone} 

For many applications to real-world observations, final-time simulation results are not sufficient. This is particularly the case for constructing observables on the past light cone, where light rays are followed from the emitting sources to the observer which requires the knowledge of the evolving space-time. 

The resulting effects along light rays, and how they relate to the relativistic coordinate choice have been extensively investigated in~\cite{Fidler:2017pnb}. Their analysis shows that in addition to the post-processing of $\xi^{\rm post}$, describing the impact of massive neutrinos, further relativistic effects need to be included on the large scales, also in the case of ordinary N-body simulations. 

Furthermore, in Eq.~(3.6) of~\cite{Adamek_2019} it was shown that general relativistic effects can be described as a modification of the lensing potential. Similar corrections can be derived for the impact of massive neutrinos, but due to the time-dependence of the metric potentials the resulting equations are more involved. 

\begin{figure}[t]
 \centering
  \includegraphics[width=0.9\linewidth]{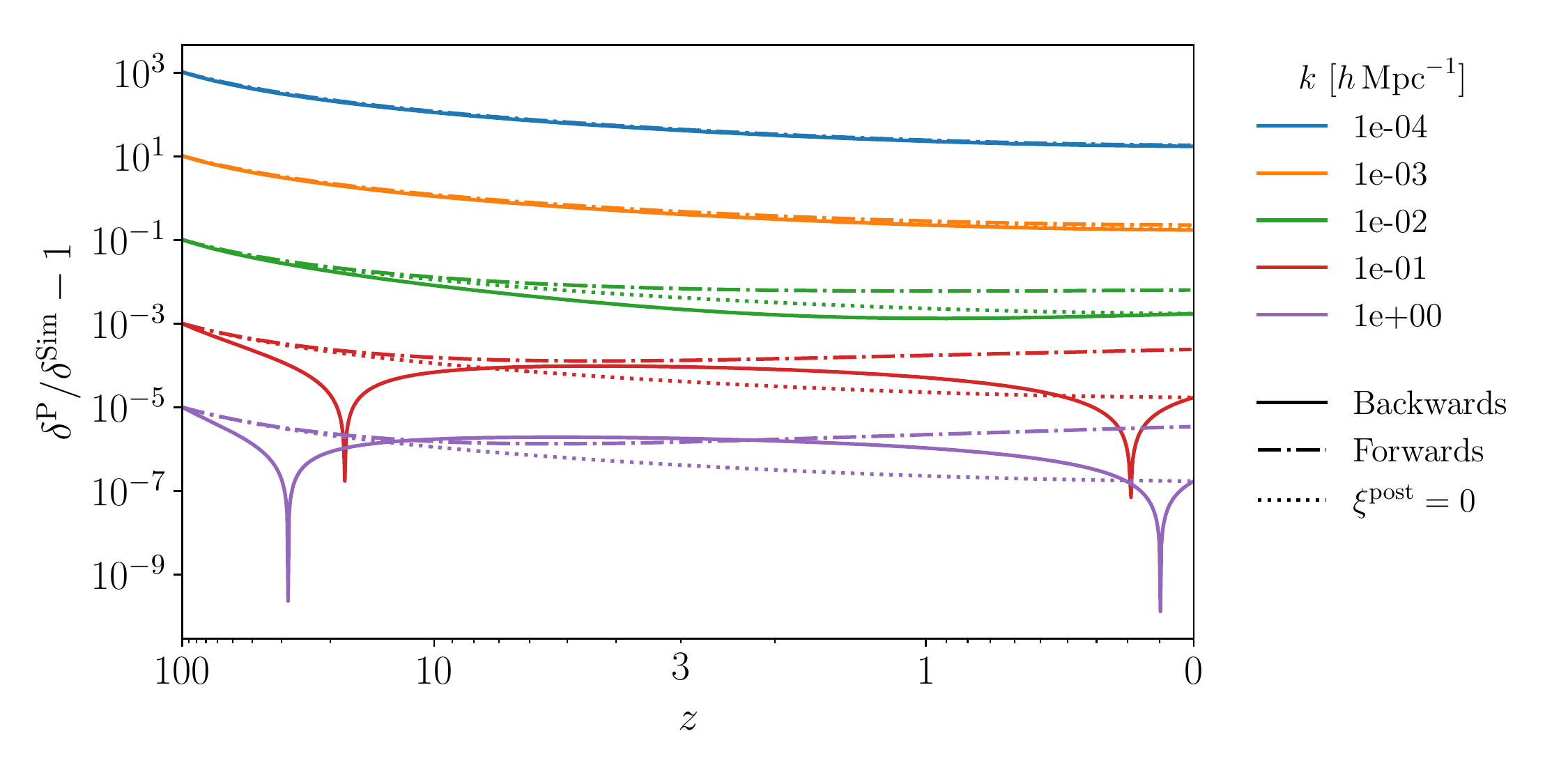}
   \caption{Shown is the relative difference between the relativistic N-boisson gauge density and the density found in a Newtonian simulation as a function of redshift. On the large scales general relativistic effects are crucial and Newtonian simulations should be interpreted using a GR dictionary. These corrections are well understood and shown in isolation in the dotted line, assuming \mbox{$\xi^{\rm post} = 0$,} implying that here we neglect the impact of massive neutrinos while keeping large scale relativistic corrections. As a comparison, the solid and dashed lines further include the impact of massive neutrinos in the forwards and backwards approach respectively. By construction, massive neutrinos in the backwards case do not require any additional post-processing at final time $z = 0$. At earlier times, the forward and backward methods both require a post-processing. However, especially at late times $z < 1$ the backwards method only introduces small mistakes if such a post-processing is omitted. Given a target precision and a range of scales and redshifts, this plot can be used to obtain the expected corrections from the post-processing. }
  \label{fig:H_T_backward}
\end{figure}

Instead of performing a full ray-tracing analysis here, we illustrate the magnitude of the resulting correction in Fig.\,\ref{fig:H_T_backward}. There we show the ratio of the relativistic N-boisson gauge density (see appendix \ref{app:Nmgauge} for details) obtained using the GR-dictionary \cite{Fidler:2017pnb} in relation to the unprocessed simulation densities for the forwards and backwards Nm~gauge as a function of redshift. We note however, as indicated above,
this includes large-scale relativistic effects independent of massive neutrinos. To highlight the impact of those in isolation we show in a dotted line the density obtained when neglecting $\xi^{\rm post}$ in the GR dictionary and thus only taking into account relativistic corrections that are also present in ordinary N-body simulations on large scales. 

We find that, as expected, the impact of gauge effects is suppressed on small scales where the Newtonian limit applies, and the Newtonian simulation can be analysed effectively without any post-processing. By contrast, large scales are subject to important corrections from GR and massive neutrinos. On intermediate scales, when the general relativistic terms start to become subdominant, we find corrections of a few percent whose amplitude directly depends on the presence of massive neutrinos. 

We note that in the backwards approach, discussed in App.\,\ref{app:backwards}, the neutrino-dependent correction $\xi^{\rm post}$ can be 
adjusted to be zero at final time. At earlier times, the correction is significantly smaller than in the forwards case. Depending on the survey and target precision, backwards approaches thus may make the neutrino dependent post-processing redundant, even at earlier times. Fig.\,\ref{fig:H_T_backward} therefore quantifies the error that occurs when a simulation in a neutrino cosmology is interpreted (erroneously) in the N-boisson gauge instead of the Nm gauge.

In summary, we have shown that, regardless of the specific method used to incorporate massive neutrinos, the gauge transformation associated to the relativistic position of the matter particles in the simulation is non-trivial on large scales, and should be taken into account.  The analysis shows that on large scales $\sim 10^{-3} h \rm Mpc^{-1}$, the combined impact of massive neutrinos and relativistic correction can be included by modifying the lensing potential as shown in~\cite{Adamek_2019} for massless neutrinos.
 By contrast, on sufficiently small scales and within the backwards approach, the impact of neutrinos on the relativistic position is negligible and thus, the simulation output can be approximately interpreted directly in the N-boisson gauge --- as it is implicitly assumed in e.g.~\cite{Tram:2018znz}.  Ideally, we would like to apply a single gauge transformation for correcting all particle positions on the ``thick'' lightcone at final time (i.e., a lightcone with a buffer, see e.g.\ \cite{Adamek:2018rru}), however we have not yet done such an analysis.

\section{Conclusions}\label{sec:conclusions}

We have performed fast neutrino simulations based on the recently introduced Newtonian motion approach. Our approach determines the effective shift $\fett{L}$ on matter particle positions due to the accumulated effect of massive neutrinos, which is added to the final particle positions in a pure matter simulation. This approach should be contrasted with complementary methods where the effect of massive neutrinos is (actively) incorporated in the Poisson solver of the simulation at subsequent time-steps (e.g., neutrino particle methods \cite{Brandbyge:2008rv, 2011MNRAS.410.1647A, 2012MNRAS.420.2551B, Villaescusa-Navarro:2013pva, Carbone:2016nzj,Brandbyge_2019, Adamek:2017uiq, Banerjee:2018bxy, Emberson:2016ecv} or grid-based methods \cite{Brandbyge:2008js,Brandbyge:2016raj,Tram:2018znz}). In our approach we assume that massive neutrinos are accurately described within linear theory, but we incorporate their feedback on matter non-perturbatively. 
The linear neutrino assumption is well justified for sufficiently light neutrinos, which we have explicitly verified by comparing our results against the simulations of Ref.\,\cite{Adamek:2017uiq} that treat neutrinos non-linearly. Specifically, Ref.\,\cite{Adamek:2017uiq} carried out simulations for the neutrino mass range  0.1\,eV$ \leq \sum m_\nu \leq 0.3$\,eV, and for the power-suppresion plot we find excellent agreement, see figure~\ref{fig:suppression_forward-comparison}. 

On a technical level, our approach assumes a weak-field description of general relativity which we have shown to be not invalidated for neutrino masses of at least $\sum m_\nu \leq 0.5$\,eV (see figure \ref{fig:H_T_forward}).
For the explicit numerical implementation of our method, only three simple modifications to a standard simulation code are necessary.
Firstly, our framework provides the initial conditions for the pure matter simulations, which amounts to initialising the particle velocities and displacements in an independent manner (as opposed to the commonly employed Zel'dovich initial conditions).
Secondly, the Friedmann equation in the Newtonian simulation code must be updated so that the Hubble expansion rate takes the effect from massive neutrinos into account. Thirdly, 
we determine the coordinate shift $\fett{L}$ from the linear Boltzmann code {\sc class} and use it to post-process the output of the simulation at final time, by shifting all particle positions accordingly (see figure~\ref{fig:displacementfield}).
Then, the final simulation output is in accordance with weak-field general relativity, and most importantly includes the dominant gravitational effects stemming from light massive neutrinos. 

Our approach is numerically as efficient as regular Newtonian N-body simulations since each of our steps in the pipeline can be performed independently. 
Because our fast method for incorporating massive neutrinos is relativistic, our simulation results naturally come with a relativistic space-time that can be
used to perform raytracing studies on the past light cone. In the present paper, we have not performed such light-cone studies but provided detailed and explicit instructions for doing so (see section~\ref{sec:lightcone}). 

One important conclusion of our work is that we have shown that the coordinate shift~$\fett{L}$ induced through the presence of massive neutrinos is suppressed on small scales. This implies that for investigating structure formation at sufficiently small scales, a simple Newtonian simulation with no post-processing suffices, provided that the change in the Hubble rate due to neutrinos is taken into account.

An interesting extension of our method would be to allow for the {\it a posteriori} change of neutrino mass, thereby opening the portal for computationally cheap parameter scans. It appears that this could be accommodated in our framework with just two additional post-processing steps:
(1) a suitably generalised coordinate transformation is required which takes the change of neutrino mass into account;
and (2) the background cosmology / Hubble rate of the simulation needs to be updated. The first step can be done within the Newtonian motion gauge framework while the latter would utilize the methods described in~\cite{Angulo:onetofitthemall}, where the set of cosmological parameters is changed after the simulation is completed by a rescaling of length, mass, velocity, time and fluctuation amplitude of a simulation snapshot. 
In a forthcoming work we will investigate such avenues in detail, and furthermore will make the required codes for generating the initial conditions  and for the post-processing publicly available.

Finally, by now a large abundance of different methods exists to evolve massive neutrinos for large-scale structure. All these methods involve different assumptions and approximations and come at largely different numerical costs. It would be timely for the community to carry out a thorough in-depth comparison of these different approaches and determine accuracy vs. efficiency in each case.

\section*{Acknowledgements}
We thank Julian Adamek and  Yvonne Y.Y.\ Wong for useful discussions, as well as Raul Angulo and Julian Adamek for valuable comments on the draft.
C.P.\ and C.F.\ thank the Laboratoire Lagrange in Nice for their hospitality and support. 
C.R.\ is a Marie Sk\l odowska–Curie Fellow and acknowledges funding from the People Programme (Marie Curie Actions) of the European Union H2020 Programme under Grant Agreement No.\ 795707 (COSMO-BLOW-UP). 
O.H.\ acknowledges funding from the European Research Council (ERC) under the European Union’s Horizon 2020 research and innovation programme (Grant Agreement No. 679145, project “COSMO-SIMS”).
Simulations were performed with computing resources granted by RWTH Aachen University under project jara0184.

\appendix

\section{Nm gauges} \label{app:Nmgauge}

For reasons of completeness, we provide in this appendix more details and definitions for our Newtonian motion gauge approach.
The Nm gauges \cite{Fidler:2015npa,Fidler:2017pnb} are the class of gauges in which dark matter particles follow Newtonian trajectories, and non-Newtonian corrections are absorbed into the choice of coordinates. Ordinary Newtonian N-body simulations can be interpreted in agreement with general relativity within the coordinates of such a gauge. 
	
We define a yet unfixed gauge by the metric  
	\begin{subequations}
    \label{metric-potentials}
    \begin{align}
    g_{00} &= -a^2(\tau) \left[ 1 + 2 A(\fett{x},\tau) \right] \,, \\
    g_{0i} &=  -a^2(\tau)  \hat\nab_i  B(\fett{x},\tau)  \,,\\
    g_{ij} &= a^2(\tau) \left[ \delta_{ij} \left( 1 + 2 \HL(\fett{x},\tau) \right) + 2 \left(  \hat\nab_i \hat\nab_j + \frac {\delta_{ij}}  {3} \Delta\right) \HT(\fett{x},\tau)  \right] \,.
    \end{align}
    \end{subequations}
    with the conformal time $\tau$, the scale factor $a=a(\tau)$, considering only scalar perturbations. 
	We further define the normalised gradient operator 
    $\hat\nab_i \equiv -(-\nab^2)^{-1/2} \nab_i$ which translates into $-\ii \hat k_i$ in Fourier space, where $\hat k_i \equiv k_i/|\bm{k}|$.
	
	In the case of a universe that is filled with only cold dark matter and a cosmological constant the N-boisson gauge with $\HT = 3 \zeta$, $\kop B = \dot{H}_{\rm T}$ is a particularly simple Nm gauge, where $\zeta$ is the gauge invariant curvature perturbation and $\kop$  the operator $(-\Delta)^{1/2}$ which corresponds to the magnitude of $k$ in Fourier space.

	Neutrinos remain well described by linear theory up to the small scales and late times~\cite{Adamek:2017uiq}, due to their large thermal velocities that hinder non-linear growth. Furthermore, on the scales where the non-linearities become relevant, the neutrino power is already strongly suppressed versus dark matter due to neutrino free streaming. Neutrino non-linearities therefore do not have an important back-reaction on the massive species. This allows us to formulate a surprisingly simple Nm gauge including massive Neutrinos. 
	Instead of simulating a shared fluid, as described in~\cite{Fidler:2018bkg} and required for example in the case of warm dark matter, we can evolve only the baryons and dark matter in the non-linear Newtonian simulation and keep the Neutrinos linear at all times and incorporate their impact at the level of the coordinate system. Several improvements to this strategy can be made, but for the currently allowed Neutrino masses the simplest idea is precise enough. 
	
	We therefore employ a single fluid Nm gauge as described in~\cite{Fidler:2017pnb}. We extend the analysis by incorporating massive neutrinos at the linear level. 
	The temporal gauge is chosen by the requirement to be consistent with weak-field gravity and is identical with the temporal gauge condition in the Poisson gauge, 
	\be
	\kop B = \dot{H}_{\rm T} \,.
	\ee
	The spatial gauge is fixed by the requirement to match the cold dark matter trajectories to the Newtonian limit (see also Sec.\ref{sec:recap}). Following the arguments presented in \cite{Fidler:2017pnb} this is equivalent to the differential equation 
	\begin{align}\nonumber
(\partial_\tau +\Hc)\dot{H}_{\rm T} &= 4\pi G a^2 \Big[\delta\rho_{\gamma} + \delta\rho_{\nu} +3 \Hc (\rho_{\gamma} + \rho_{\nu} + p_{\gamma}+ p_{\nu})\kop^{-1} (v - \kop^{-1}\dot{H}_{\rm T}) \\
&- (\rho_{\rm cdm}+\rho_{\rm b})(3\zeta -  \HT)\Big] + 8 \pi G a^2 \Sigma\,,
 \label{gaugedefinition}
\end{align}
	where $\rho_{\gamma}$ and $\rho_{\nu}$ are the photon and neutrino densities, $\rho_{\rm cdm}$ the dark matter ones and $\rho_{\rm b}$ describes the baryons. All terms in this equation are small in the weak-field limit and suppressed on the non-linear scales. Therefore, a linear Boltzmann solver such as \textsc{class} or {\sc{camb}} is able to generate a valid solution for the metric potential $\HT$. 
	
Using this gauge condition, the relativistic Euler equation for cold dark matter takes a Newtonian form. However, the relativistic continuity equation still does not match the Newtonian one realised in a N-body simulation. This is simply due to the fact that the relativistic densities incorporate volume perturbations, while the Newtonian ones are simple counting densities. We therefore identify the density in the Newtonian simulation with the relativistic counting density:
\be 
 \rho^{\rm Nm}_{\rm m} =  (1 - 3\HL)\rho_{\rm m}^{\rm N} \,.
\ee
From these two corrections we can construct a GR dictionary for a Newtonian simulation. An ordinary Newtonian N-body simulation can then be employed to describe the cold species ($\rho_{\rm cdm}$ and $\rho_{\rm b}$) while the impact of massive neutrinos (and photons) is incorporated after the simulation is completed by interpreting the output on the given Nm gauge.

\section{More details about the backwards method}\label{app:backwards}

Being a second-order differential equation, Eq.\,\eqref{gaugedefinition} shows that the Nm gauges are not completely fixed and a 2-parameter residual gauge freedom remains. This gauge freedom can be related to the freedom to choose the initial density and velocity in the Newtonian simulation. 
The most natural gauge fixing is to start on the N-boisson gauge ($\HT = 3\zeta$ and $\dot{H}_{\rm T} = 3\dot{\zeta}$). This will lead to a description where we slowly drift away from the N-boisson gauge and the post-processing $\xi^{\rm post}$ grows towards the later times. 

Alternatively we can choose to find different gauge conditions in which we are closer to the N-boisson gauge at the final time and not initially, to remove the necessity to do a post-processing. The naive idea is to demand that we finish exactly on the N-boisson gauge, leading to the back-scaling method in a massless neutrino cosmology.

We find that another choice is to be preferred: We fix the gauge by demanding that $\HT = 3\zeta$ at both the initial and final time, not demanding anything from the derivatives. This means that we are neither on the N-boisson gauge at the initial time, nor at the final time. However, in both cases we are relatively close, with particles being at the N-boisson gauge position, but having a different set of velocities (due to mismatches in $\dot{H}_{\rm T}$). This choice turns out to be numerically more stable and provides values of $\HT$ that remain reasonably close to $3\zeta$ at later times such that the post-processing may be neglected given a certain target precision. 

We numerically create the solution for $\HT$, and the corresponding initial conditions for the N-body simulation, using a 'shooting' method. We perform a suite of forward runs in \CLASS{}~with varying initial conditions and choose the one that matches our demands the best. We then iteratively optimise around the best solution. 

Note that these initial conditions have different properties than classical back-scaling; in particular they cannot be trivially generated from the present day cold matter power spectrum by solving the evolution backwards. They especially do not depend on the present day total matter power-spectrum, which is not a suitable way to initialise neutrino simulations and leads to significant mistakes in the subsequent non-linear evolution; see Ref.\,\cite{Fidler:2018stc} for more details.

\bibliography{references}{}

\providecommand{\href}[2]{#2}\begingroup\raggedright\begin{thebibliography}{10}

\bibitem{euclid}
{\scshape EUCLID} collaboration, R.~Laureijs et~al., \emph{{Euclid Definition
  Study Report}},  \href{https://arxiv.org/abs/1110.3193}{{\tt 1110.3193}}.

\bibitem{lsst}
{\scshape LSST Science, LSST Project} collaboration, P.~A. Abell et~al.,
  \emph{{LSST Science Book, Version 2.0}},
  \href{https://arxiv.org/abs/0912.0201}{{\tt 0912.0201}}.

\bibitem{Hearin_2012}
A.~P. Hearin, A.~R. Zentner and Z.~Ma, \emph{{General Requirements on Matter
  Power Spectrum Predictions for Cosmology with Weak Lensing Tomography}},
  \href{http://dx.doi.org/10.1088/1475-7516/2012/04/034}{\emph{JCAP} {\bf 1204}
  (2012) 034}, [\href{https://arxiv.org/abs/1111.0052}{{\tt 1111.0052}}].

\bibitem{Lesgourgues:2012uu}
J.~Lesgourgues and S.~Pastor, \emph{{Neutrino mass from Cosmology}},
  \href{http://dx.doi.org/10.1155/2012/608515}{\emph{Adv. High Energy Phys.}
  {\bf 2012} (2012) 608515}, [\href{https://arxiv.org/abs/1212.6154}{{\tt
  1212.6154}}].

\bibitem{Jimenez:2016ckl}
R.~Jimenez, C.~P. Garay and L.~Verde, \emph{{Neutrino footprint in Large Scale
  Structure}}, \href{http://dx.doi.org/10.1016/j.dark.2016.11.004}{\emph{Phys.
  Dark Univ.} {\bf 15} (2017) 31--34},
  [\href{https://arxiv.org/abs/1602.08430}{{\tt 1602.08430}}].

\bibitem{Brandbyge:2008rv}
J.~Brandbyge, S.~Hannestad, T.~Haugbølle and B.~Thomsen, \emph{{The Effect of
  Thermal Neutrino Motion on the Non-linear Cosmological Matter Power
  Spectrum}},
  \href{http://dx.doi.org/10.1088/1475-7516/2008/08/020}{\emph{JCAP} {\bf 0808}
  (2008) 020}, [\href{https://arxiv.org/abs/0802.3700}{{\tt 0802.3700}}].

\bibitem{2011MNRAS.410.1647A}
S.~Agarwal and H.~A. Feldman, \emph{{The effect of massive neutrinos on the
  matter power spectrum}},
  \href{http://dx.doi.org/10.1111/j.1365-2966.2010.17546.x}{\emph{Mon. Not.
  Roy. Astron. Soc.} {\bf 410} (2011) 1647},
  [\href{https://arxiv.org/abs/1006.0689}{{\tt 1006.0689}}].

\bibitem{2012MNRAS.420.2551B}
S.~Bird, M.~Viel and M.~G. Haehnelt, \emph{{Massive Neutrinos and the
  Non-linear Matter Power Spectrum}},
  \href{http://dx.doi.org/10.1111/j.1365-2966.2011.20222.x}{\emph{Mon. Not.
  Roy. Astron. Soc.} {\bf 420} (2012) 2551--2561},
  [\href{https://arxiv.org/abs/1109.4416}{{\tt 1109.4416}}].

\bibitem{Villaescusa-Navarro:2013pva}
F.~Villaescusa-Navarro, F.~Marulli, M.~Viel, E.~Branchini, E.~Castorina,
  E.~Sefusatti et~al., \emph{{Cosmology with massive neutrinos I: towards a
  realistic modeling of the relation between matter, haloes and galaxies}},
  \href{http://dx.doi.org/10.1088/1475-7516/2014/03/011}{\emph{JCAP} {\bf 1403}
  (2014) 011}, [\href{https://arxiv.org/abs/1311.0866}{{\tt 1311.0866}}].

\bibitem{Carbone:2016nzj}
C.~Carbone, M.~Petkova and K.~Dolag, \emph{{DEMNUni: ISW, Rees-Sciama, and
  weak-lensing in the presence of massive neutrinos}},
  \href{http://dx.doi.org/10.1088/1475-7516/2016/07/034}{\emph{JCAP} {\bf 1607}
  (2016) 034}, [\href{https://arxiv.org/abs/1605.02024}{{\tt 1605.02024}}].

\bibitem{Brandbyge_2019}
J.~Brandbyge, S.~Hannestad and T.~Tram, \emph{{Momentum space sampling of
  neutrinos in $N$-body simulations}},
  \href{http://dx.doi.org/10.1088/1475-7516/2019/03/047}{\emph{JCAP} {\bf 1903}
  (2019) 047}, [\href{https://arxiv.org/abs/1806.05874}{{\tt 1806.05874}}].

\bibitem{Adamek:2017uiq}
J.~Adamek, R.~Durrer and M.~Kunz, \emph{{Relativistic N-body simulations with
  massive neutrinos}},
  \href{http://dx.doi.org/10.1088/1475-7516/2017/11/004}{\emph{JCAP} {\bf 1711}
  (2017) 004}, [\href{https://arxiv.org/abs/1707.06938}{{\tt 1707.06938}}].

\bibitem{Banerjee:2018bxy}
A.~Banerjee, D.~Powell, T.~Abel and F.~Villaescusa-Navarro, \emph{{Reducing
  Noise in Cosmological N-body Simulations with Neutrinos}},
  \href{http://dx.doi.org/10.1088/1475-7516/2018/09/028}{\emph{JCAP} {\bf 1809}
  (2018) 028}, [\href{https://arxiv.org/abs/1801.03906}{{\tt 1801.03906}}].

\bibitem{Emberson:2016ecv}
J.~D. Emberson et~al., \emph{{Cosmological neutrino simulations at extreme
  scale}}, \href{http://dx.doi.org/10.1088/1674-4527/17/8/85}{\emph{Res.
  Astron. Astrophys.} {\bf 17} (2017) 085},
  [\href{https://arxiv.org/abs/1611.01545}{{\tt 1611.01545}}].

\bibitem{Ichiki_2012}
K.~Ichiki and M.~Takada, \emph{{The impact of massive neutrinos on the
  abundance of massive clusters}},
  \href{http://dx.doi.org/10.1103/PhysRevD.85.063521}{\emph{Phys. Rev.} {\bf
  D85} (2012) 063521}, [\href{https://arxiv.org/abs/1108.4688}{{\tt
  1108.4688}}].

\bibitem{2014PhRvD..89f3502L}
M.~{LoVerde} and M.~{Zaldarriaga}, \emph{{Neutrino clustering around spherical
  dark matter halos}},
  \href{http://dx.doi.org/10.1103/PhysRevD.89.063502}{\emph{Phys. Rev.} {\bf
  D89} (Mar, 2014) 063502}, [\href{https://arxiv.org/abs/1310.6459}{{\tt
  1310.6459}}].

\bibitem{2014PhRvD..90h3518L}
M.~{LoVerde}, \emph{{Spherical collapse in
  {\ensuremath{\nu}}{\ensuremath{\Lambda}}CDM}},
  \href{http://dx.doi.org/10.1103/PhysRevD.90.083518}{\emph{Phys. Rev.} {\bf
  D90} (Oct, 2014) 083518}, [\href{https://arxiv.org/abs/1405.4858}{{\tt
  1405.4858}}].

\bibitem{Aghanim:2018eyx}
{\scshape Planck} collaboration, N.~Aghanim et~al., \emph{{Planck 2018 results.
  VI. Cosmological parameters}},  \href{https://arxiv.org/abs/1807.06209}{{\tt
  1807.06209}}.

\bibitem{Vagnozzi_2017}
S.~Vagnozzi, E.~Giusarma, O.~Mena, K.~Freese, M.~Gerbino, S.~Ho et~al.,
  \emph{{Unveiling $\nu$ secrets with cosmological data: neutrino masses and
  mass hierarchy}},
  \href{http://dx.doi.org/10.1103/PhysRevD.96.123503}{\emph{Phys. Rev.} {\bf
  D96} (2017) 123503}, [\href{https://arxiv.org/abs/1701.08172}{{\tt
  1701.08172}}].

\bibitem{Brandbyge:2008js}
J.~Brandbyge and S.~Hannestad, \emph{{Grid Based Linear Neutrino Perturbations
  in Cosmological N-body Simulations}},
  \href{http://dx.doi.org/10.1088/1475-7516/2009/05/002}{\emph{JCAP} {\bf 0905}
  (2009) 002}, [\href{https://arxiv.org/abs/0812.3149}{{\tt 0812.3149}}].

\bibitem{Brandbyge:2016raj}
J.~Brandbyge, C.~Rampf, T.~Tram, F.~Leclercq, C.~Fidler and S.~Hannestad,
  \emph{{Cosmological $N$-body simulations including radiation perturbations}},
  \href{http://dx.doi.org/10.1093/mnrasl/slw235}{\emph{Mon. Not. Roy. Astron.
  Soc.} {\bf 466} (2017) L68--L72},
  [\href{https://arxiv.org/abs/1610.04236}{{\tt 1610.04236}}].

\bibitem{Tram:2018znz}
T.~Tram, J.~Brandbyge, J.~Dakin and S.~Hannestad, \emph{{Fully relativistic
  treatment of light neutrinos in $N$-body simulations}},
  \href{http://dx.doi.org/10.1088/1475-7516/2019/03/022}{\emph{JCAP} {\bf 1903}
  (2019) 022}, [\href{https://arxiv.org/abs/1811.00904}{{\tt 1811.00904}}].

\bibitem{2013MNRAS.428.3375A}
Y.~Ali-Haimoud and S.~Bird, \emph{{An efficient implementation of massive
  neutrinos in non-linear structure formation simulations}},
  \href{http://dx.doi.org/10.1093/mnras/sts286}{\emph{Mon. Not. Roy. Astron.
  Soc.} {\bf 428} (2012) 3375--3389},
  [\href{https://arxiv.org/abs/1209.0461}{{\tt 1209.0461}}].

\bibitem{Liu:2017now}
J.~Liu, S.~Bird, J.~M.~Z. Matilla, J.~C. Hill, Z.~Haiman, M.~S. Madhavacheril
  et~al., \emph{{MassiveNuS: Cosmological Massive Neutrino Simulations}},
  \href{http://dx.doi.org/10.1088/1475-7516/2018/03/049}{\emph{JCAP} {\bf 1803}
  (2018) 049}, [\href{https://arxiv.org/abs/1711.10524}{{\tt 1711.10524}}].

\bibitem{Brandbyge:2009ce}
J.~Brandbyge and S.~Hannestad, \emph{{Resolving Cosmic Neutrino Structure: A
  Hybrid Neutrino N-body Scheme}},
  \href{http://dx.doi.org/10.1088/1475-7516/2010/01/021}{\emph{JCAP} {\bf 1001}
  (2010) 021}, [\href{https://arxiv.org/abs/0908.1969}{{\tt 0908.1969}}].

\bibitem{Bird:2018all}
S.~Bird, Y.~Ali-Haïmoud, Y.~Feng and J.~Liu, \emph{{An Efficient and Accurate
  Hybrid Method for Simulating Non-Linear Neutrino Structure}},
  \href{http://dx.doi.org/10.1093/mnras/sty2376}{\emph{Mon. Not. Roy. Astron.
  Soc.} {\bf 481} (2018) 1486--1500},
  [\href{https://arxiv.org/abs/1803.09854}{{\tt 1803.09854}}].

\bibitem{Zennaro:2019aoi}
M.~Zennaro, R.~E. Angulo, G.~Aricò, S.~Contreras and M.~Pellejero-Ibáñez,
  \emph{{How to add massive neutrinos to your $\Lambda$CDM simulation -
  extending cosmology rescaling algorithms}},
  \href{http://dx.doi.org/10.1093/mnras/stz2612}{\emph{Mon. Not. Roy. Astron.
  Soc.} {\bf 489} (2019) 5938--5951},
  [\href{https://arxiv.org/abs/1905.08696}{{\tt 1905.08696}}].

\bibitem{Zennaro:2016nqo}
M.~Zennaro, J.~Bel, F.~Villaescusa-Navarro, C.~Carbone, E.~Sefusatti and
  L.~Guzzo, \emph{{Initial Conditions for Accurate N-Body Simulations of
  Massive Neutrino Cosmologies}},
  \href{http://dx.doi.org/10.1093/mnras/stw3340}{\emph{Mon. Not. Roy. Astron.
  Soc.} {\bf 466} (2017) 3244--3258},
  [\href{https://arxiv.org/abs/1605.05283}{{\tt 1605.05283}}].

\bibitem{Angulo:onetofitthemall}
R.~E. Angulo and S.~D.~M. White, \emph{{One simulation to fit them all -
  changing the background parameters of a cosmological N-body simulation}},
  \href{http://dx.doi.org/10.1111/j.1365-2966.2010.16459.x}{\emph{Mon. Not.
  Roy. Astron. Soc.} {\bf 405} (2010) 143},
  [\href{https://arxiv.org/abs/0912.4277}{{\tt 0912.4277}}].

\bibitem{Wong:2008ws}
Y.~Y.~Y. Wong, \emph{{Higher order corrections to the large scale matter power
  spectrum in the presence of massive neutrinos}},
  \href{http://dx.doi.org/10.1088/1475-7516/2008/10/035}{\emph{JCAP} {\bf 0810}
  (2008) 035}, [\href{https://arxiv.org/abs/0809.0693}{{\tt 0809.0693}}].

\bibitem{Blas:2014hya}
D.~Blas, M.~Garny, T.~Konstandin and J.~Lesgourgues, \emph{{Structure formation
  with massive neutrinos: going beyond linear theory}},
  \href{http://dx.doi.org/10.1088/1475-7516/2014/11/039}{\emph{JCAP} {\bf 1411}
  (2014) 039}, [\href{https://arxiv.org/abs/1408.2995}{{\tt 1408.2995}}].

\bibitem{Fuhrer:2014zka}
F.~Führer and Y.~Y.~Y. Wong, \emph{{Higher-order massive neutrino
  perturbations in large-scale structure}},
  \href{http://dx.doi.org/10.1088/1475-7516/2015/03/046}{\emph{JCAP} {\bf 1503}
  (2015) 046}, [\href{https://arxiv.org/abs/1412.2764}{{\tt 1412.2764}}].

\bibitem{Senatore:2017hyk}
L.~Senatore and M.~Zaldarriaga, \emph{{The Effective Field Theory of
  Large-Scale Structure in the presence of Massive Neutrinos}},
  \href{https://arxiv.org/abs/1707.04698}{{\tt 1707.04698}}.

\bibitem{deBelsunce:2018xtd}
R.~de~Belsunce and L.~Senatore, \emph{{Tree-Level Bispectrum in the Effective
  Field Theory of Large-Scale Structure extended to Massive Neutrinos}},
  \href{http://dx.doi.org/10.1088/1475-7516/2019/02/038}{\emph{JCAP} {\bf 1902}
  (2019) 038}, [\href{https://arxiv.org/abs/1804.06849}{{\tt 1804.06849}}].

\bibitem{Saito_2008}
S.~Saito, M.~Takada and A.~Taruya, \emph{{Impact of massive neutrinos on
  nonlinear matter power spectrum}},
  \href{http://dx.doi.org/10.1103/PhysRevLett.100.191301}{\emph{Phys. Rev.
  Lett.} {\bf 100} (2008) 191301}, [\href{https://arxiv.org/abs/0801.0607}{{\tt
  0801.0607}}].

\bibitem{Chisari:2011iq}
N.~E. Chisari and M.~Zaldarriaga, \emph{{Connection between Newtonian
  simulations and general relativity}},
  \href{http://dx.doi.org/10.1103/PhysRevD.84.089901,
  10.1103/PhysRevD.83.123505}{\emph{Phys. Rev.} {\bf D83} (2011) 123505},
  [\href{https://arxiv.org/abs/1101.3555}{{\tt 1101.3555}}].

\bibitem{Fidler:2015npa}
C.~Fidler, C.~Rampf, T.~Tram, R.~Crittenden, K.~Koyama and D.~Wands,
  \emph{{General relativistic corrections to $N$-body simulations and the
  Zel'dovich approximation}},
  \href{http://dx.doi.org/10.1103/PhysRevD.92.123517}{\emph{Phys. Rev.} {\bf
  D92} (2015) 123517}, [\href{https://arxiv.org/abs/1505.04756}{{\tt
  1505.04756}}].

\bibitem{Hahn:2016roq}
O.~Hahn and A.~Paranjape, \emph{{General relativistic screening in cosmological
  simulations}},
  \href{http://dx.doi.org/10.1103/PhysRevD.94.083511}{\emph{Phys. Rev.} {\bf
  D94} (2016) 083511}, [\href{https://arxiv.org/abs/1602.07699}{{\tt
  1602.07699}}].

\bibitem{Fidler:2017pnb}
C.~Fidler, T.~Tram, C.~Rampf, R.~Crittenden, K.~Koyama and D.~Wands,
  \emph{{General relativistic weak-field limit and Newtonian N-body
  simulations}},
  \href{http://dx.doi.org/10.1088/1475-7516/2017/12/022}{\emph{JCAP} {\bf 1712}
  (2017) 022}, [\href{https://arxiv.org/abs/1708.07769}{{\tt 1708.07769}}].

\bibitem{Fidler:2016tir}
C.~Fidler, T.~Tram, C.~Rampf, R.~Crittenden, K.~Koyama and D.~Wands,
  \emph{{Relativistic Interpretation of Newtonian Simulations for Cosmic
  Structure Formation}},
  \href{http://dx.doi.org/10.1088/1475-7516/2016/09/031}{\emph{JCAP} {\bf 1609}
  (2016) 031}, [\href{https://arxiv.org/abs/1606.05588}{{\tt 1606.05588}}].

\bibitem{Fidler:2018bkg}
C.~Fidler, A.~Kleinjohann, T.~Tram, C.~Rampf and K.~Koyama, \emph{{A new
  approach to cosmological structure formation with massive neutrinos}},
  \href{http://dx.doi.org/10.1088/1475-7516/2019/01/025}{\emph{JCAP} {\bf 1901}
  (2019) 025}, [\href{https://arxiv.org/abs/1807.03701}{{\tt 1807.03701}}].

\bibitem{Fidler:2018stc}
C.~Fidler and A.~Kleinjohann, \emph{{Suitable Initial Conditions for Newtonian
  Simulations with Massive Neutrinos}},
  \href{http://dx.doi.org/10.1088/1475-7516/2019/06/018}{\emph{JCAP} {\bf 1906}
  (2019) 018}, [\href{https://arxiv.org/abs/1810.12019}{{\tt 1810.12019}}].

\bibitem{class2}
D.~Blas, J.~Lesgourgues and T.~Tram, \emph{{The Cosmic Linear Anisotropy
  Solving System (CLASS) II: Approximation schemes}},
  \href{http://dx.doi.org/10.1088/1475-7516/2011/07/034}{\emph{JCAP} {\bf 1107}
  (2011) 034}, [\href{https://arxiv.org/abs/1104.2933}{{\tt 1104.2933}}].

\bibitem{Lewis:1999bs}
A.~Lewis, A.~Challinor and A.~Lasenby, \emph{{Efficient computation of CMB
  anisotropies in closed FRW models}},
  \href{http://dx.doi.org/10.1086/309179}{\emph{Astrophys. J.} {\bf 538} (2000)
  473--476}, [\href{https://arxiv.org/abs/astro-ph/9911177}{{\tt
  astro-ph/9911177}}].

\bibitem{Adamek:2015eda}
J.~Adamek, D.~Daverio, R.~Durrer and M.~Kunz, \emph{{General relativity and
  cosmic structure formation}},
  \href{http://dx.doi.org/10.1038/nphys3673}{\emph{Nature Phys.} {\bf 12}
  (2016) 346--349}, [\href{https://arxiv.org/abs/1509.01699}{{\tt
  1509.01699}}].

\bibitem{Adamek:2016zes}
J.~Adamek, D.~Daverio, R.~Durrer and M.~Kunz, \emph{{gevolution: a cosmological
  N-body code based on General Relativity}},
  \href{http://dx.doi.org/10.1088/1475-7516/2016/07/053}{\emph{JCAP} {\bf 1607}
  (2016) 053}, [\href{https://arxiv.org/abs/1604.06065}{{\tt 1604.06065}}].

\bibitem{Springel:2005mi}
V.~Springel, \emph{{The Cosmological simulation code GADGET-2}},
  \href{http://dx.doi.org/10.1111/j.1365-2966.2005.09655.x}{\emph{Mon. Not.
  Roy. Astron. Soc.} {\bf 364} (2005) 1105--1134},
  [\href{https://arxiv.org/abs/astro-ph/0505010}{{\tt astro-ph/0505010}}].

\bibitem{Ma:1995ey}
C.-P. Ma and E.~Bertschinger, \emph{{Cosmological perturbation theory in the
  synchronous and conformal Newtonian gauges}},
  \href{http://dx.doi.org/10.1086/176550}{\emph{Astrophys. J.} {\bf 455} (1995)
  7--25}, [\href{https://arxiv.org/abs/astro-ph/9506072}{{\tt
  astro-ph/9506072}}].

\bibitem{Dakin_2019}
J.~Dakin, J.~Brandbyge, S.~Hannestad, T.~HaugbØlle and T.~Tram,
  \emph{nuconcept: cosmological neutrino simulations from the non-linear
  boltzmann hierarchy},
  \href{http://dx.doi.org/10.1088/1475-7516/2019/02/052}{\emph{JCAP} {\bf 2019}
  (Feb, 2019) 052–052}.

\bibitem{Adamek:2017grt}
J.~Adamek, J.~Brandbyge, C.~Fidler, S.~Hannestad, C.~Rampf and T.~Tram,
  \emph{{The effect of early radiation in N-body simulations of cosmic
  structure formation}},
  \href{http://dx.doi.org/10.1093/mnras/stx1157}{\emph{Mon. Not. Roy. Astron.
  Soc.} {\bf 470} (2017) 303--313},
  [\href{https://arxiv.org/abs/1703.08585}{{\tt 1703.08585}}].

\bibitem{Hannestad:2020rzl}
S.~Hannestad, A.~Upadhye and Y.~Y. Wong, \emph{{Spoon or slide? The non-linear
  matter power spectrum in the presence of massive neutrinos}},
  \href{https://arxiv.org/abs/2006.04995}{{\tt 2006.04995}}.

\bibitem{Fidler:2017ebh}
C.~Fidler, T.~Tram, C.~Rampf, R.~Crittenden, K.~Koyama and D.~Wands,
  \emph{{Relativistic initial conditions for N-body simulations}},
  \href{http://dx.doi.org/10.1088/1475-7516/2017/06/043}{\emph{JCAP} {\bf 1706}
  (2017) 043}, [\href{https://arxiv.org/abs/1702.03221}{{\tt 1702.03221}}].

\bibitem{Adamek_2019}
J.~Adamek and C.~Fidler, \emph{The large-scale general-relativistic correction
  for newtonian mocks},
  \href{http://dx.doi.org/10.1088/1475-7516/2019/09/026}{\emph{JCAP} {\bf 2019}
  (Sep, 2019) 026–026}.

\bibitem{Adamek:2018rru}
J.~Adamek, C.~Clarkson, L.~Coates, R.~Durrer and M.~Kunz, \emph{{Bias and
  scatter in the Hubble diagram from cosmological large-scale structure}},
  \href{http://dx.doi.org/10.1103/PhysRevD.100.021301}{\emph{Phys. Rev. D} {\bf
  100} (2019) 021301}, [\href{https://arxiv.org/abs/1812.04336}{{\tt
  1812.04336}}].

\end{thebibliography}\endgroup
\bibliographystyle{JHEP}

\end{document}